\documentclass[useAMS,usenatbib]{mn2e}
\usepackage{hyperref}
\usepackage{epsfig}
\usepackage{epstopdf}
\usepackage{graphicx}
\usepackage{amssymb}
\usepackage{amsmath}
\usepackage{graphicx}
\usepackage{subcaption}
\usepackage[utf8]{inputenc}
\usepackage[export]{adjustbox}
\usepackage{wrapfig}
\usepackage{textcomp}
\usepackage[usenames,dvipsnames]{color}
\usepackage{gensymb}
\usepackage{booktabs}
\usepackage{amsmath}
\usepackage{lscape}
\usepackage[english]{babel}
\usepackage{graphicx}
\usepackage{float}
\usepackage{fixltx2e}
\usepackage{soul}
\usepackage[normalem]{ulem}
\usepackage{ulem}

\DeclareMathAlphabet{\mathbbold}{U}{bbold}{m}{n}
\title[Photo-$z$ in COSMOS and XMM-LSS]{Hybrid photometric redshifts for sources in the COSMOS and XMM-LSS fields}
\author[Peter Hatfield]{P. W. Hatfield$^{1}$\thanks{peter.hatfield@physics.ox.ac.uk}, M. J. Jarvis$^{1,2}$, N.Adams$^{3}$, R.A.A. Bowler$^{1,3}$, B.H\"au\ss ler$^{4}$, K. J. Duncan$^{5}$\\
$^{1}$Astrophysics, University of Oxford, Denys Wilkinson Building, Keble Road, Oxford, OX1 3RH, UK\\
$^{2}$Department of Physics, University of the Western Cape, Bellville 7535, South Africa\\
$^{3}$Jodrell Bank Centre for Astrophysics, University of Manchester, Oxford Road, Manchester, M13 9PL, UK\\
$^{4}$European Southern Observatory, Alonso de Cordova 3107, Vitacura, Casilla 19001, Santiago, Chile\\
$^{5}$SUPA, Institute for Astronomy, Royal Observatory, Blackford Hill, Edinburgh, EH9 3HJ, UK\\
}

\begin{document}

\pagerange{\pageref{firstpage}--\pageref{lastpage}} \pubyear{2021}

\maketitle

\label{firstpage}

\begin{abstract}

In this paper we present photometric redshifts for 2.7 million galaxies in the XMM-LSS and COSMOS fields, both with rich optical and near-infrared data from VISTA and HyperSuprimeCam. Both template fitting (using galaxy and Active Galactic Nuclei templates within LePhare) and machine learning (using GPz) methods are run on the aperture photometry of sources selected in the $K_{\mathrm{s}}$-band. The resulting predictions are then combined using a Hierarchical Bayesian model, to produce consensus photometric redshift point estimates and probability distribution functions that outperform each method individually. Our point estimates have a root mean square error of $\sim0.08-0.09$, and an outlier fraction of $\sim3-4$ percent when compared to spectroscopic redshifts. We also compare our results to the COSMOS2020 photometric redshifts, which contains fewer sources, but had access to a larger number of bands and greater wavelength coverage, finding that comparable photo-$z$ quality can be achieved (for bright and intermediate luminosity sources where a direct comparison can be made). Our resulting redshifts represent the most accurate set of photometric redshifts (for a catalogue this large) for these deep multi-square degree multi-wavelength fields to date.

\end{abstract}


\section{Introduction}

Many contemporary astronomical studies in extragalactic astrophysics and cosmology involve estimating the redshifts of large numbers of distant sources (typically galaxies). Galaxy redshift estimates are necessary  to probe the time evolution of the Universe, as well as to correctly calculate galaxy properties - estimates of absolute luminosity and related measurements such as galaxy stellar mass rely on the estimated redshift being correct (\citealp{Hsieh2014}).

Galaxy, and also Active Galactic Nuclei (AGN), redshifts can be calculated from their electromagnetic spectrum in two main ways, from spectroscopy or from photometry. \textit{Spectroscopic redshifts} (``spec-$z$'s'') are calculated by detecting a known spectral (normally emission) line or feature with a spectrograph, and measuring the `shift'  from the known rest frame wavelength/frequency. \textit{Photometric redshifts} (``photo-$z$'s'') are calculated by measuring the brightness of the source in $N$ broad wavelength ranges, and making a redshift prediction based on this coarse spectral data. Spectroscopic redshifts are far more precise than photo-$z$'s (as long as the spectral feature is correctly identified) but are more costly (in terms of telescope time), so are generally restricted to much smaller samples. Spectroscopic and photometric redshift measurements are thus appropriate for different science goals (\citealp{Fernandez-Soto2001}).

There are two main methods for photometric redshift calculation (e.g. \citealp{Salvato2019}); `template fitting' and `machine learning'. Template fitting methods typically use a number of galaxy or AGN template spectra (either empirical or synthetic), and use a $\chi^2$-minimisation-like method to find the `best' redshift estimate. Template-fitting based codes in regular use include Photometric Analysis for Redshift Estimate (LePhare, \citealp{Arnouts1999,Ilbert2006}), Bayesian Photometric Redshifts (BPZ, \citealp{Benitez2000,Benitez2004,Coe2006}), the Zurich Extragalactic Bayesian Redshift Analyzer (ZEBRA, \citealp{Feldmann2006}), EAzY (\citealp{Brammer2008}) and Phosphoros (Paltani et al. in prep).

Machine learning photo-$z$ methods take a highly empirical approach. The prediction task is treated as a supervised machine learning problem, where predictions must be made based on the photometry, and galaxies with known (usually spectroscopic) redshifts are used as labelled training data. Widely used machine learning photo-$z$ codes include Artificial Neural Network Redshifts (ANNz2, \citealp{Collister2004,Sadeh2016}), Trees for photo-$z$ (TPZ, \citealp{CarrascoKind2013}), Self Organizing Map Redshifts (SOMz, \citealp{CarrascoKind2013}), Machine-learning Estimation Tool for Accurate PHOtometric Redshifts (METAPHOR, \citealp{Cavuoti2017}), and many more.

Template-fitting and machine learning methods typically only make identical predictions in the simplest of cases. In general the methods make different predictions, with differing claims of levels of precision achieved, and giving different redshift probability distribution functions (pdfs), with corresponding advantages and disadvantages for different science cases (outlined in \citealp{Salvato2019}). Each method is typically reliable in different parts of colour-magnitude space. This presents an opportunity to achieve redshift prediction performance beyond that of each method individually, seeking `the best of both worlds'. This has been demonstrated with a number of different approaches, for a number of different data sets (\citealp{Brodwin2006,Carrasco2014,Duncan2018b,Schmidt2020,Hatfield2020a}).\textcolor{white}{XXX XXX XXX XXX XXX XXX XXX XXX XXX XXX XXX XXX XXX XXX XXX XXX XXX XXX XXX XXX XXX XXX XXX XXX XXX XXX XXX XXX XXX XXX XXX XXX XXX XXX XXX XXX XXX XXX XXX XXX XXX XXX XXX XXX XXX XXX XXX XXX XXX XXX XXX XXX XXX XXX XXX XXX XXX XXX XXX XXX XXX XXX XXX XXX XXX XXX XXX XXX XXX XX XXX XXX}

In this work we present photometric redshift calculations that seek to achieve `the best of both worlds' for the rich multi-wavelength data sets that span the COSMOS and XMM-LSS fields, two of the most well-studied extragalactic fields. These redshifts will be key for a large range of extragalactic studies in these fields.

The structure of this paper is as follows. In Section 2 we describe the data used in this study. In Section 3 we describe the algorithms used, namely GPz (\citealp{Almosallam2016a,Almosallam2016b}) and {\sc LePhare} (\citealp{Arnouts1999,Ilbert2006}). In Section 4 we discuss our results, we discuss the significance in Section 5, and we conclude in Section 6. AB magnitudes are used throughout (\citealp{Oke1983}).

\section{Data} \label{sec:data}

\begin{figure*}
\includegraphics[scale=0.5]{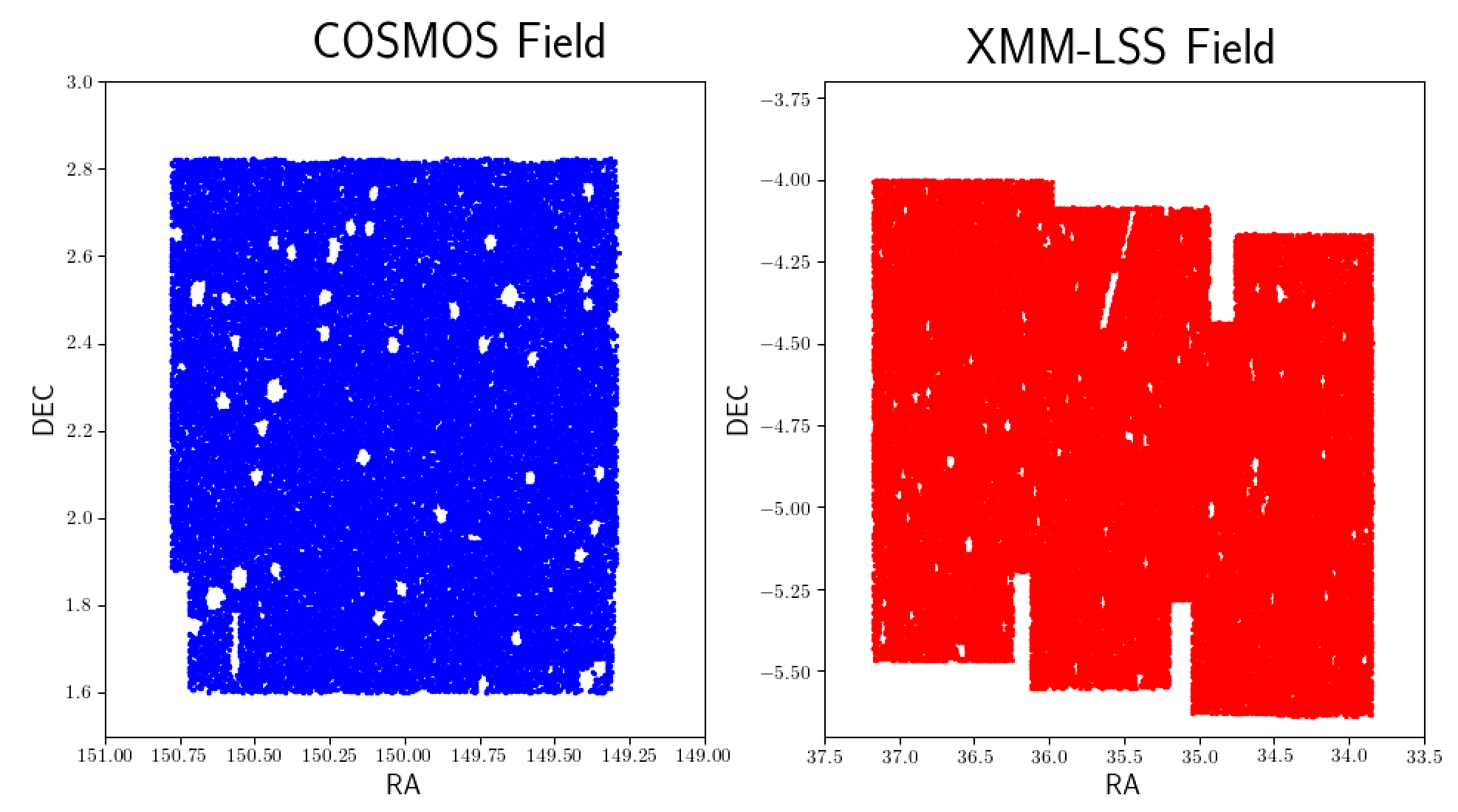}
\caption{Field geometry of the source populations used in this study, spanning the COSMOS and XMM-LSS fields. Different scales are used for the two sub-plots. Holes and gaps in the fields are from bright nearby stars and artefacts in the imaging.}
\label{fig:field_geometry}
\end{figure*}

The data in this work covers the COSMOS and XMM-Newton Large-Scale Structure (XMM-LSS) fields - see Figure \ref{fig:field_geometry}. These fields represent two of the deepest and widest fields used in extragalactic high-redshift survey astronomy, regularly used for a large number of wide-ranging studies e.g. \citet{Frayer2009,Darvish2017a, Ata2021}, \citet{Pacaud2007, Clerc2014,Chen2018,Hale2018}. The catalogues we  use are described in \citet{Bowler2020} and \citet{Adams2020}, which, in order to ensure consistency, used identical procedures to extract the photometry across the two fields. The data is thus very homogenous across the two fields. Sources were selected in the $K_s$ band (down to a limiting magnitude of $K_{s}=24.8$ in COSMOS and $K_{s}=23.9$ in XMM-LSS), and forced photometry was performed on all the other bands. 2'' diameter circular apertures were used, which had an aperture correction applied by a model generated with PSFEx (\citealp{Bertin2011}) for each band. In the COSMOS field 995,049 sources were identified, with 1,674,689 sources identified in the XMM-LSS field (2,669,738 in total).

The photometry used spans 10 filters;  $u$ (Canada-France-Hawaii telescope Large Area U-band Deep Survey, CLAUDS, for both COSMOS and the wider area XMM-LSS, \citealp{Sawicki2019}),  $grizy$ (Hyper Suprime-Cam Subaru Strategic Program, HSC-SSP, for both COSMOS and XMM-LSS, \citealp{Aihara2017a}, \citealp{Kawanomoto2018}) and YJHK$_{\mathrm{s}}$  (Visible and Infrared Survey Telescope for Astronomy, VISTA, VISTA Deep Extragalactic Observations, VIDEO for XMM-LSS, \citealp{Jarvis2013} and UltraVISTA for COSMOS, \citealp{McCracken2012}). The photometric depths in the COSMOS (XMM-LSS\footnote{Depths are not quite identical for the three VISTA tiles in XMM-LSS; here the deepest data value is quoted, see Table 1 in \citet{Adams2020} for more details.}) field are $u=27.0$ ($=26.9$), $g$$=27.2$ ($=27.0$),  $r$$=26.8$ ($=26.5$),  $i$$=26.6$ ($=26.4$),  $z$$=25.9$ ($=26.3$), $y_{\textrm{HSC}}$$=25.5$ ($=25.6$),  $Y_{\textrm{VISTA}}=25.5$ ($=25.2$), $J=25.3$ ($=24.7$), $H=25.0$ ($=24.3$) and $K_{\mathrm{s}}=24.8$ ($=23.9$).

The fields have spectroscopic redshifts (used in the training process for the machine learning based photometric redshifts) from a range of sources\footnote{This spectroscopic data set is constructed largely similarly to the Catalog of Spectroscopic Redshifts from the Hyper Suprime-Cam Subaru Strategic Program Public Data Release, \url{https://hsc-release.mtk.nao.ac.jp/doc/index.php/dr1_specz/}}. The spectroscopic redshifts are taken from the VVDS (\citealp{LeFevre2013}), VANDELS (\citealp{McLure2018}; \citealp{Pentericci2018}), Z-COSMOS (\citealp{Lilly2009}), SDSS-DR12 (\citealp{Alam2015}), 3D-HST (\citealp{Skelton2014}; \citealp{Momcheva2016}), Primus (\citealp{Coil2011}; \citealp{Cool2013}), DEIMOS-10K (\citealp{Hasinger2018}) and FMOS (\citealp{Silverman2015}) surveys. There were 25,268 spectroscopic redshifts in the COSMOS field, and 14,846 in the XMM-LSS field.

We would note that machine learning based photo-$z$ methods are reliant on the accuracy of the spectroscopic redshifts in the training sample. If the spectroscopic redshifts used in the training process are inaccurate then machine learning methods will simply reproduce the incorrect values (see for example \citealp{Stylianou2022}). For this reason we only used the most secure spectroscopic redshifts that have flags indicating high quality (confidence of $\geq$ 95 per cent). Where a source had a secure spectroscopic redshift available from more than one survey, the mean of the secure redshifts was used. Furthermore as discussed in \citet{Hatfield2020a} we found that the Primus spectroscopic redshifts could be inconsistent with the higher-resolution spectroscopic redshifts at $z>1$. For this reason we only use the $z<1$ Primus spectroscopic redshifts.

\section{Algorithms} \label{sec:Algorithms}

\subsection{Template Fitting: LePhare} \label{sec:lephare}

Our template-based photo-$z$'s are calculated in a very similar manner to \citet{Adams2020}, using {\sc LePhare} (\citealp{Arnouts1999,Ilbert2006}). The only difference is that rather than using AGN templates to find a $\chi^2_{\mathrm{AGN}}$ to later use for classification/contamination control, we instead run {\sc LePhare} twice, once with galaxy templates, and once with AGN templates, to obtain two template fitting based photo-$z$ pdfs.

When finding a galaxy template-fitting photo-$z$ pdf the COSMOS SED template set (\citealp{Ilbert2009}) was used, where 32 templates were sourced from \citet{Polletta2007} with the GRASIL code (\citealp{Silva1998}) and from \citet{Bruzual2003}. The templates cover a range of galaxy morphologies and spectral types (E, S0, Sa, Sb, Sc, Sd, Sdm) and have rest-frame wavelength ranges that cover our optical and near-infrared dataset. Within the fitting process, each of these templates is allowed to be modified for the effects of dust attenuation using the \citet{Calzetti2000} attenuation law and an attenuation value in the range E(B-V)=0-1.5. At each redshift, we use the \citet{Madau1995} treatment for absorption by the intergalactic medium (IGM). For the AGN template-fitting photo-$z$ pdf calculation, spectra for AGN from \citet{Salvato2009} were instead fit. Zero-point corrections to the photometry were made as in \citet{Adams2020}. We do not address the potential impact of AGN variability on photo-$z$ quality (e.g. \citealp{Simm2015}) in this work. For the {\sc LePhare} calculations broad uniform priors over absolute magnitude ($-28<M_{\mathrm{abs}}<-10$), redshift ($0<z<9$), and dust attenuation ($0<$E(B-V)$<1.5$), were used in this work, with an informative prior over redshift being introduced at the Hierarchical Bayesian Combination stage, see Section \ref{sec:HBC}. The fiducial cosmology used was is a standard Flat cosmology of $\Omega_{\mathrm{M}}=0.3$, $\Omega_{\mathrm{\Lambda}}=0.7$ and $H_0=70$ km s$^{-1}$ Mpc$^{-1}$.

\subsection{Machine Learning: GPz} \label{sec:gpz}

GPz is a supervised machine learning algorithm developed for the problem of calculating photometric redshifts (\citealp{Almosallam2016a,Almosallam2016b}). The algorithm is `sparse Gaussian process' (GP) based, e.g. see \citet{Rasmussen2006}. The input data for the algorithm consists of sources with photometry and spectroscopic redshifts (the `labels'). The algorithm is trained on this data, and then makes predictions for data with no spectroscopic redshift.

GPz has been used on a number of data sets e.g. \citet{Gomes2017,Duncan2018b, Zuntz2021}, but (as with all algorithms), has some deficiencies, in particular 1) making poor predictions in parts of parameter space underrepresented in the training data and 2) only producing Gaussian pdfs (where in fact non-Gaussian, multi-modal pdfs might typically better represent our uncertainty). \citet{Hatfield2020a} (see also \citealp{Duncan2022}) investigated a number of ways in which the bias introduced by these issues could be reduced. They found a combination (referred to as `GMM-All') of 1) reweighting validation data to be closer to the target data, 2) dividing up the colour-magnitude space into regions and modelling each one separately, and 3) resampling the data many times based on the uncertainties on the photometry improved the resulting predictions. For the data set under consideration here, we calculate GPz pdfs using GMM-All, rather than using the base GPz pdfs. Missing bands and uncertainties on photometry were treated by adding the noise variance to the basis functions and constructing a joint distribution of the input parameters as per Section 5 of \citet{Almosallam2017}. \textcolor{white}{XXX XXX XXX XXX XXX XXX XXX XXX XXX XXX XXX XXX XXX XXX XXX XXX XXX XXX XXX XXX XXX XXX XXX XXX XXX XXX XXX XXX XXX XXX XXX XXX XXX XXX XXX XXX XXX XXX XXX XXX XXX XXX XXX XXX XXX XXX XXX XXX XXX XXX XXX XXX XXX XXX XXX XXX XXX XXX XXX XXX XXX XXX XXX XXX XXX XXX XXX XXX XXX XX XXX XXX}

Unless otherwise stated, we use the GPz settings in \autoref{table-settings} (see \citealp{Almosallam2016a,Almosallam2016b} for precise definitions and interpretations).

 \begin{table*}
\caption{Parameter setting of GPz.}
\begin{center}
\begin{tabular}{| l | l | l |}
    Parameter 	&	Value		&	Description\\	\hline
	$m$			&	500		&	Number of basis functions; complexity of GP, in general higher $m$ is more accurate but longer run time\\
	maxIter		&	500		&	Maximum number of iterations\\
	maxAttempts	&	50		&	Maximum iterations to attempt if there is no progress on the validation set\\
	method		&	GPVC	&	Bespoke covariances on each basis function\\
	normalize	&	True	&	Pre-process the input by subtracting the means and dividing by the standard deviations\\
	joint		&	True	& Jointly learn a prior linear mean-function\\
  \end{tabular}
\end{center}
\label{table-settings}
\end{table*}

\subsection{Hierarchical Bayesian Combination} \label{sec:HBC}

We use a Hierarchical Bayesian (HB) model similar to that described in \citet{Duncan2018a} and \citet{Duncan2018b} (which builds on \citealp{Dahlen2013}).This method seeks to combine the pdfs from $n$ different redshift estimates\footnote{The $n$ photo-$z$ estimates could be different methods within the same `class' of estimator, or could be different classes of estimate completely. \citet{Duncan2018a} for example combine three different template based photo-$z$ estimates ($n=3$). In \citet{Duncan2018b} an ML based photo-$z$ and a template based photo-$z$ are combined ($n=2$). In principle in future other independent approaches to redshift estimation could be incorporated, for example cluster-$z$ (e.g. \citealp{Rahman2015}) and photo-geometric redshifts (e.g. \citealp{Sonnenfeld2021}).} to achieve a consensus pdf that is more accurate than the individual estimators. Note this method is different to that described in section 5.1 of \citet{Hatfield2020a}, which combined template and ML methods by accepting the ML prediction in the interpolative regime, and the template prediction in the extrapolative regime. The \citet{Hatfield2020a} approach permitted use of knowledge of where we would expect the machine learning predictions to be reliable, and where we would expect it to be unreliable, but didn't enable the full information in the individual pdfs to be used. An alternative method of combining pdfs to that described in \citet{Duncan2018a,Duncan2018b} is using a Fr\'echet mean method, as described in \citet{Kodra2019}.

When calculating consensus pdfs, we first used a Hierarchical Bayesian model to combine the galaxy and AGN template pdfs ($n=2$), to produce a `best' template-fitting redshift estimate pdf. This `best' template-fitting redshift estimate pdf is then combined with the machine learning pdf with a Hierarchical Bayesian model (again $n=2$).  See \citet{Duncan2018a} for a full description of the method, but for each source and each redshift estimate $i$ we define:

\begin{equation}
\begin{aligned}
P(z,f_{\mathrm{bad}})_{i}=P(z|\mathrm{bad} \; \mathrm{measurement})_{i} f_{\mathrm{bad}} +\\
P(z|\mathrm{good} \; \mathrm{measurement})_{i} (1-f_{\mathrm{bad}})
\end{aligned}
\end{equation}

where $f_{\mathrm{bad}}$ is a parameter describing the probability of an estimate being incorrect,  $P(z|\mathrm{good} \; \mathrm{measurement})_{i} $ is the probability distribution assuming that the estimator is correct (i.e. equal to the probability distribution from the estimator), $P(z|\mathrm{bad} \; \mathrm{measurement})_{i} $ is the probability distribution assuming that the estimator is incorrect (typically chosen as an appropriate prior), and $i$ indexes the $n$ methods (so $i$ indexes over 1 and 2 for the AGN and the galaxy template fits for the first Hierarchical Bayesian Model, and then over 1 and 2 again for the template-based and the ML based pdfs for the second Hierarchical Bayesian Model). $P(z,f_{\mathrm{bad}})_{i}$ thus describes a pdf from that estimator, allowing for the possibility of the estimate being incorrect.

These $n$ distributions are then combined in the following way:
\begin{equation} \label{eq:beta}
P(z,f_{\mathrm{bad}})= \prod^{n}_{i=1} P(z,f_{\mathrm{bad}})_{i}^{1/\beta_{i}} , 
\end{equation}

where the $\beta_{i}$ are constants that encode the weights and covariances between the different measurements. Equation \ref{eq:beta} represents a small generalisation over \citet{Duncan2018a}, for which each estimate had the same $\beta$ values. The product iterates over the $n$ different photo-$z$ pdfs being combined, which are indexed by $i$.

Finally $f_{\mathrm{bad}}$ is marginalised over to get a final pdf:
\begin{equation}
P(z)= \int^{f_{\mathrm{bad}}^{\mathrm{max}} }_{f_{\mathrm{bad}}^{\mathrm{min}} }   P(z,f_{\mathrm{bad}}) \mathrm{d}f_{\mathrm{bad}} ,
\end{equation}

where $f_{\mathrm{bad}}^{\mathrm{min}} $ and $f_{\mathrm{bad}}^{\mathrm{max}} $ are constants representing the minimum and maximum of the range marginalised over.

The choice of $\beta_{i}$ characterises the weighting and degree of correlation between estimates. When all the $\beta_{i}$ are chosen to be equal (to some $\beta$), the two extremes are $\beta=1$ and $\beta=\frac{1}{n}$. $\beta=1$ corresponds to simply multiplying the pdfs together i.e. treating them as completely independent and multiplying the probabilities. $\beta=\frac{1}{n}$ corresponds to taking the geometric mean of the estimates for the case where they are fully covariant (based completely on the same underlying data). These extremes can to some degree be thought of as corresponding to `AND' ($\beta=1$, all the estimates are giving independent information that should be incorporated into the prediction), and `OR' ($\beta=\frac{1}{n}$, not all the predictions can be independently true) (\citealp{Duncan2018a}).

\subsection{Combining Galaxy and AGN Template Based PDFs} \label{sec:gal_agn_combination_description}

Our galaxy and AGN template-based pdfs are highly covariant in that they are based on the same data (the photometry) - the only difference is the modelling. If we were agnostic between the galaxy and AGN pdfs, $\beta=\frac{1}{2}$ would be a natural choice (as $n=2$). However for individual galaxies we are not agnostic - we can use the $\chi^2$ to indicate which of the galaxy and AGN templates were better fitting. We hence allow the $\beta_{i}$ to vary, requiring that $\sum \frac{1}{\beta_{i}}=1$ (essentially generalising the geometric mean to the weighted geometric mean). We choose
\begin{equation}
\frac{1}{\beta_{\mathrm{galaxy}}}=\frac{\exp(-\chi^{2}_{\mathrm{galaxy}}/2) }{\exp(-\chi^{2}_{\mathrm{galaxy}}/2) + \exp(-\chi^{2}_{\mathrm{AGN}}/2)}
\end{equation}
and
\begin{equation}
 \frac{1}{\beta_{\mathrm{AGN}}}=\frac{\exp(-\chi^{2}_{\mathrm{AGN}}/2) }{\exp(-\chi^{2}_{\mathrm{galaxy}}/2) + \exp(-\chi^{2}_{\mathrm{AGN}}/2)}
\end{equation}
 to reflect the probabilities implied by the $\chi^2$ values. Note that when $\chi^{2}_{\mathrm{galaxy}}=\chi^{2}_{\mathrm{AGN}}$ we recover $\beta_{i}=\frac{1}{n}=\frac{1}{2}$, and in the limit of $\chi^{2}_{\mathrm{galaxy}}/\chi^{2}_{\mathrm{AGN}}=0$ (high confidence that the source is a galaxy not an AGN) we find $ \frac{1}{\beta_{\mathrm{galaxy}}}=1$ and $ \frac{1}{\beta_{\mathrm{AGN}}}=0$ (and vice versa). We also note that one could treat all individual galaxy and AGN templates in this way and obtain a final pdf. However for the purposes of this paper we consider the simpler combination, noting that one could also in principle impose a prior based on the expected number of galaxies and AGN at any given epoch i.e. the luminosity function - however as many luminosity functions are based on photometric redshifts one risks circular arguments.

\subsection{Combining Template Based and Machine Learning Based PDFs}

\begin{figure}
\includegraphics[scale=0.45]{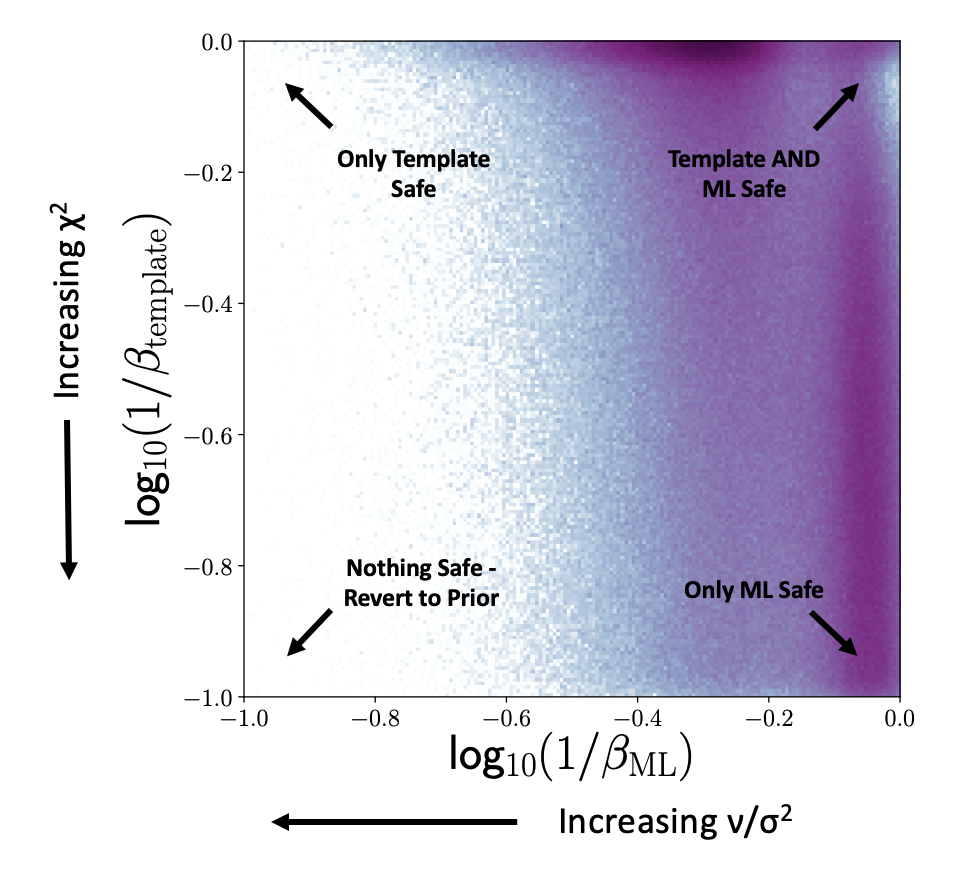}
\caption{A simple diagram illustrating which photometric redshift methods are more reliable in different parts of colour magnitude space when the template and the ML pdfs are combined (with the actual $\beta_{i}$ values from the data plotted). When the template models fit well and the ML is interpolating, both the template and the ML pdfs are reliable. When the template models fit well, but the ML is extrapolating, only the template pdf is reliable. When the template models become invalid, but the ML still has adequate training data in that part of parameter space, then the ML pdf but not the template pdf is reliable. When the template models do not fit, and the ML is extrapolating, neither approach is reliable, and the best that can be done is to revert to some broad prior.}
\label{fig:param_space}
\end{figure}

\begin{figure*}
\includegraphics[scale=0.9]{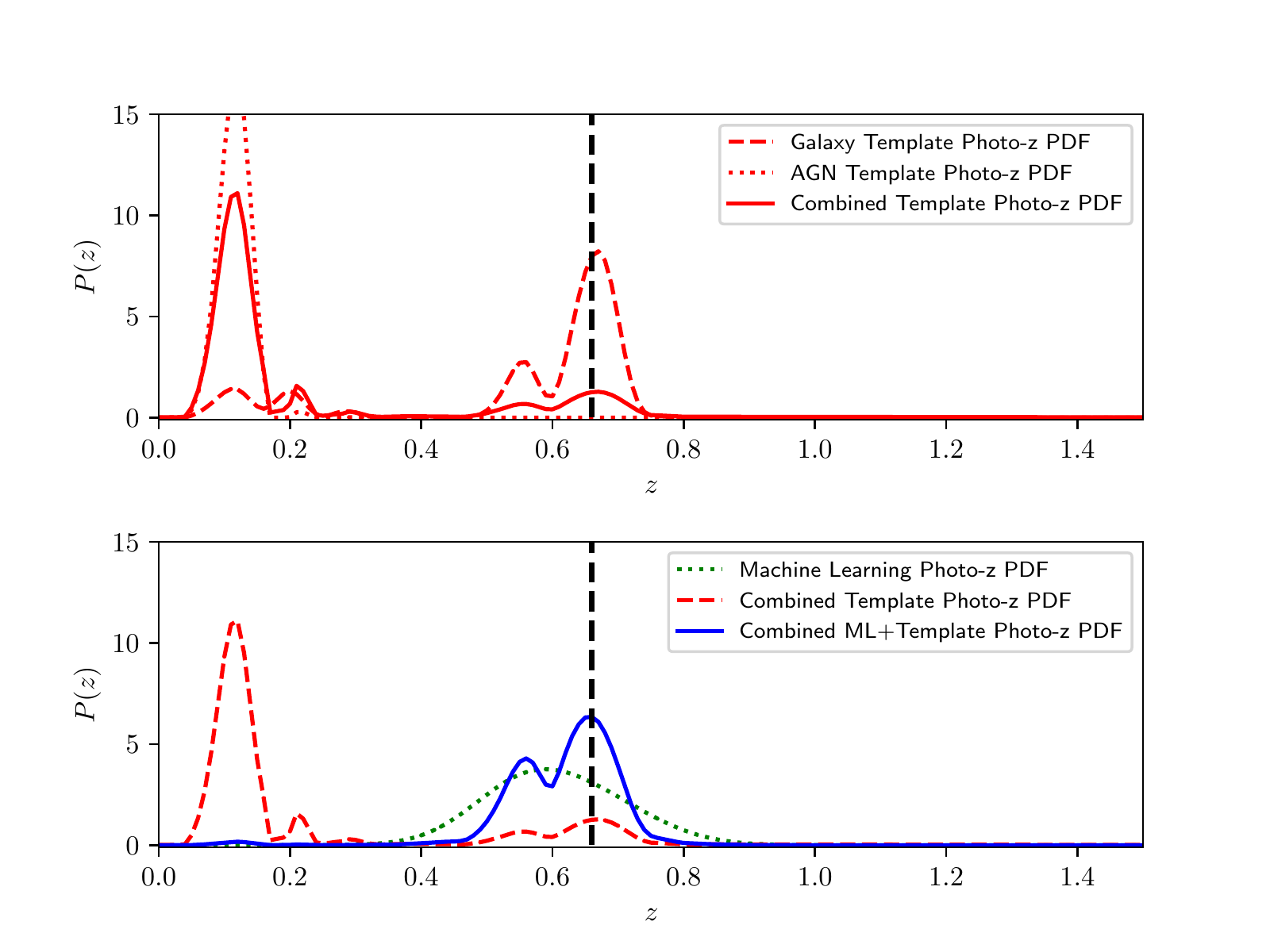}
\caption{Illustrative example of how the Hierarchical Bayesian Model used combines pdfs. The top panel shows the combination of the galaxy and AGN photo-$z$ template pdfs into a consensus template fitting pdf. The lower panel shows the combination of the consensus template fitting pdf with the ML pdf. The vertical line shows the spectroscopic redshift.}
\label{fig:Hierarchical_bayes}
\end{figure*}

When combining the template-fitting pdf and the ML pdf ($n=2$), one choice would be to use $\beta_{i}=1$ as the estimates are highly independent (the template fitting method does not have access to the information contained within the spectroscopic training data, and the ML method does not have access to our knowledge of the physics implicit within the templates)\footnote{They are not completely independent, as they both use the same photometry, so scatter on the magnitudes for the two estimates are correlated. However uncertainty on photometry is a sub-dominant source of photo-$z$ uncertainty.}. However we do have estimates of the `reliability' of the two methods (as opposed to simply the associated uncertainties). Template fitting predictions, uncertainties and pdfs are likely to be unreliable when the $\chi^2$ of the best fitting template and redshift is high e.g. the photometry is not well fit by any template. Similarly, as discussed in \citet{Hatfield2020a}, the ML predictions become less reliable in the extrapolative regime, which can be quantified by how much of the uncertainty is due to lack of data in that part of parameter space compared with the total uncertainty (which includes uncertainty from the photometry, as well as intrinsic scatter in output redshift). Thus we use $1/\beta_{\mathrm{template}}=\exp(-\frac{\chi^2}{2})$ and $1/\beta_{\mathrm{ML}}=1-\frac{\nu}{\sigma^2}$ (where $\nu$ is the variance from the lack of data using GPz, $\sigma^2$ is the total GPz variance, see \citealp{Almosallam2016b} and section 5.1 of \citealp{Hatfield2020a}). In the limit of both the ML and template-fitting being reliable the $1/\beta_{i}\rightarrow1$, in the limit of templates fitting well but extrapolating far from the spectroscopic training data $1/\beta_{\mathrm{template}}\rightarrow1$, $1/\beta_{\mathrm{ML}}\rightarrow0$, in the limit of no template fitting well, but there being sufficient training data in that part of parameter space $1/\beta_{\mathrm{template}}\rightarrow0$, $1/\beta_{\mathrm{ML}}\rightarrow1$, and finally where no template fits well and there is no nearby training data $1/\beta_{\mathrm{template}}\rightarrow0$, $1/\beta_{\mathrm{ML}}\rightarrow0$ and the resulting pdf reverts to the prior. In Figure \ref{fig:param_space} we show a schematic for how the pdfs are combined differently in different parts of parameter space, as well as how our sources actually cover the space. Note that most galaxies have at least one reliable photo-$z$ method. Figure \ref{fig:Hierarchical_bayes} shows the template and ML based pdfs for a sample galaxy, and the pdfs that result in the Hierarchical Bayesian combination. 

\begin{figure*}
\includegraphics[scale=0.6]{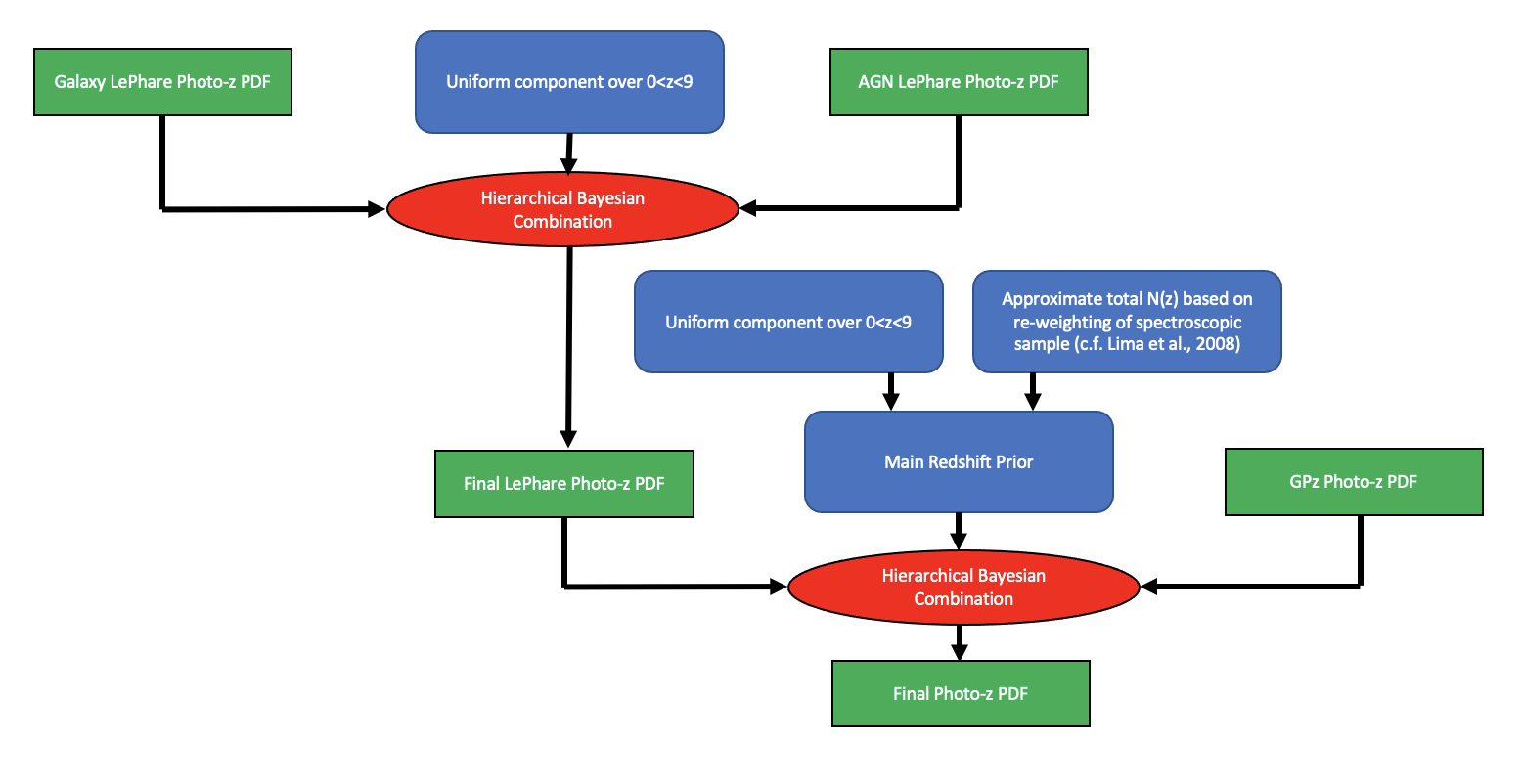}
\caption{Simple flow chart schematic to illustrate how our final pdfs are constructed. First the galaxy and AGN {\sc LePhare} template-based pdfs are combined with a Hierarchical Bayesian model. Then the resulting pdf is combined with the GPz ML pdf with a second hierarchical Bayesian model (using a prior that is a weighted combination of a uniform distribution and the approximate sample $N(z)$). }
\label{fig:directed_graph}
\end{figure*}

We use $f_{\mathrm{bad}}^{\mathrm{min}}=0 $ and $f_{\mathrm{bad}}^{\mathrm{max}}=0.05$ as per \citet{Duncan2018a}. In other words, we do not select a value for $f_{\mathrm{bad}}$ itself, but instead select the range over which it is marginalised. $f_{\mathrm{bad}}^{\mathrm{min}} $ and $f_{\mathrm{bad}}^{\mathrm{max}}$ could be chosen by a tuning process (considered in \citealp{Duncan2018a}), although this presents difficulties due to the differences between the training and test data, so there is no guarantee that any optimal values found would actually be optimal for the target data. The values from \citet{Duncan2018a} were partially based on a tuning process, but also based on typical outlier fractions for their data (and the surveys considered here have comparable properties and outlier fractions). Both we and \citet{Duncan2018a} found that extremely high values of $f_{\mathrm{bad}}^{\mathrm{max}}$ (e.g. $f_{\mathrm{bad}}^{\mathrm{max}}\geq0.5$) gave worse predictions, but fine tuning beyond choosing a value of approximately 0.05/a value comparable to the expected outlier fraction made negligible difference. Finally for $P(z|\mathrm{bad} \; \mathrm{measurement})_{i} $ we use a linear combination of a uniform distribution in redshift over $(0,9)$ (weight 0.001), and the implied approximate sample redshift distribution (weight 0.999). The approximate sample redshift distribution is constructed by taking the redshift distribution of the spectroscopic sample, where each source is weighted by how many sources of similar colour-magnitudes there are in the full sample (based on the Gaussian Mixture Models from the GMM-All calculation), similarly to as in \citet{Lima2008}\footnote{Note that this choice of prior is very similar to the `trainZ' estimator considered in \citet{Schmidt2020} - a simple estimator that assigned each source an identical redshift pdf, that of the population as a whole.}. The addition of the uniform prior is necessary to avoid the distribution being zero for sparse (mainly higher) redshifts. If not included, the prior becomes zero (as opposed to just very small) for redshifts higher than the highest spectroscopic redshift in the sample, which is unphysical (and thus even if the template estimate was highly secure at a higher redshift, the zero-weight prior would dominate and lead to a lower redshift being assigned). Results are relatively insensitive to exact choice of relative weighting of the two components of the prior and were chosen to approximately reflect how many high redshift sources might be expected, to an order of magnitude. Finally, see Figure \ref{fig:directed_graph} for a simple diagram illustrating how the two Hierarchical Bayesian models used here are connected.

\subsection{Selection of `Best' Redshift Estimate} \label{sec:best}

There is normally no single `best' choice of point estimate of redshift from a (generally imperfect) redshift pdf\footnote{See for example discussion in \citet{Duncan2019}}, as the statistical properties required depend on science goal e.g. sensitivity to outliers, redshift range of interest etc. Here we quote the mode of the pdf as the `best' estimate of redshift, although other point-estimates (e.g. median, mean) can be readily calculated from the pdfs.

\section{Results} \label{sec:corr_functions}

In this section we apply the methods discussed in Section \ref{sec:Algorithms} to the data described in Section \ref{sec:data}. To test the quality of our calculations, one approach is to compare predictions to the spectroscopic sample.  However because not all galaxies have a spectroscopic redshift, the comparisons that include $z_{\mathrm{spec}}$ represent a biased sub-set of the whole dataset. Furthermore the ML based predictions were trained on this sample, so we would expect these predictions to be much better than for unseen data (even of the same colour-magnitude distribution). Good performance on the spectroscopic sample is thus necessary but not sufficient.

In this section for the calculation of metrics we remove from the sample sources in the stellar locus (that are likely to be stars) as defined in \citet{Jarvis2013} (which follows the approach of \citealp{Baldry2010}). We also remove sources with $\chi^2_{\mathrm{Star}}<\mathrm{min}(\chi^2_{\mathrm{QSO}},\chi^2_{\mathrm{Galaxy}})$. This reduces the COSMOS and XMM-LSS samples to 815,673 and 1,557,392 respectively. The sources in the stellar locus are still assigned photometric redshifts for the released catalogue e.g. in case the stellar classification is incorrect due to scatter.

In addition to the spectroscopic sample, our photo-$z$ calculations can also be compared to the COSMOS2020 photo-$z$ catalogue of \citet{Weaver2021} (an update to the COSMOS2015 redshifts of \citealp{Laigle2016}). We use the LePhare redshifts based on the `Classic' catalogue. This data set includes optical and NIR data of similar bands to those used in this dataset over the COSMOS field. However in addition to broad-band photometry, it also used a number of medium- and narrow-band filters for the calculation of template-based photometric redshifts (36 bands used in total). Thus the COSMOS2020 photo-$z$ calculations represent an intermediate category between spectroscopic redshifts and the photo-$z$ we have calculated, in the sense that they are likely a) more accurate than our template photo-$z$, but less accurate than spectroscopic redshifts, and b) less numerous/deep than our sample, but more numerous/deep than the spectroscopic redshift sample. However, they are not used in the training process, and have a different colour-magnitude distribution to the training spectroscopic data, so represent a more realistic test of the photo-$z$ quality. To extract the COSMOS2020 photo-$z$, we cross-match to our COSMOS data (1 arcsecond max error). COSMOS2020 photo-$z$'s were found for 664,322 of our 995,049 COSMOS sources ($\sim$65 percent).

\begin{figure*}
\includegraphics[scale=0.7]{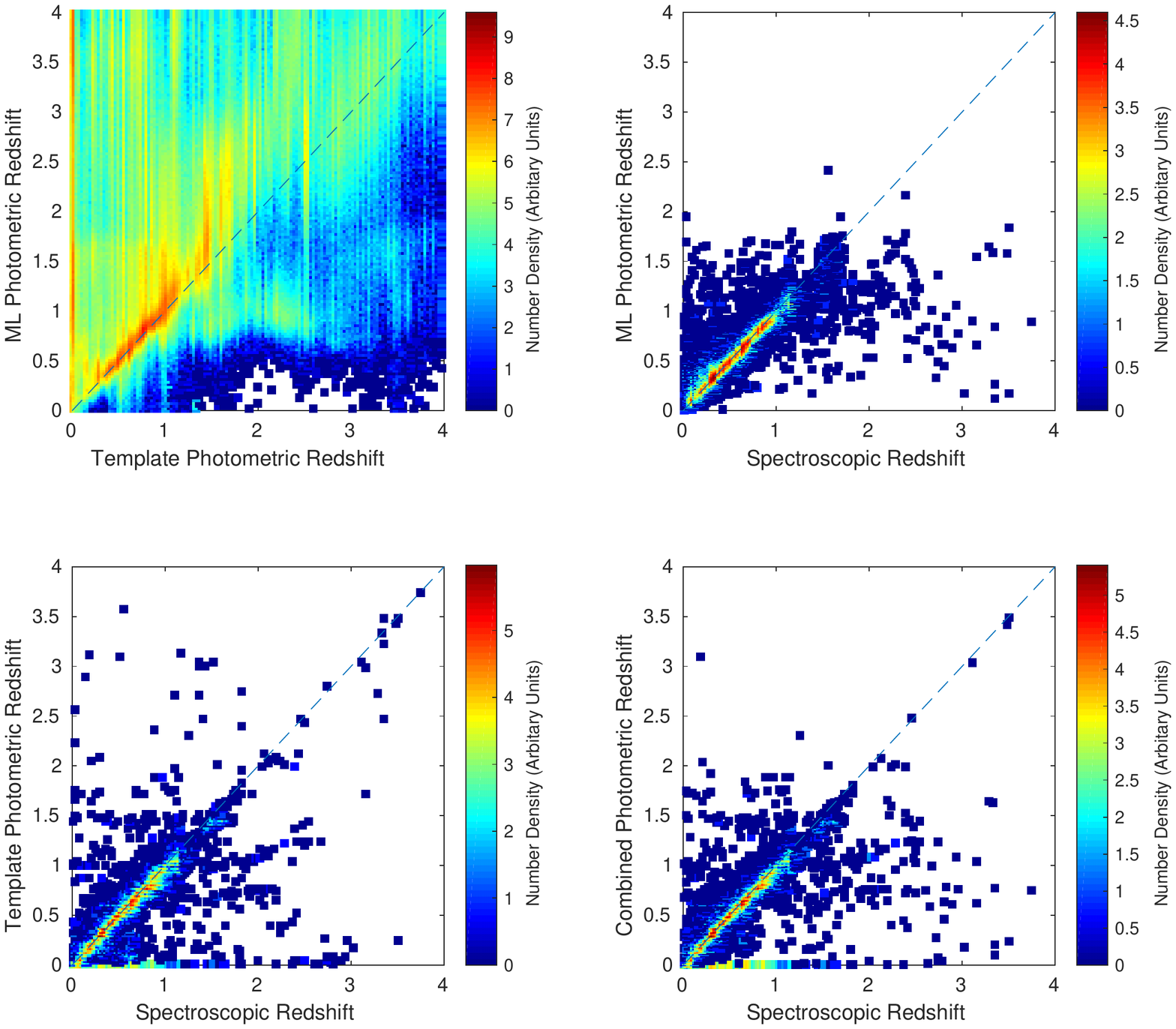}
\caption{Comparisons of the $z_{\mathrm{phot}}$ (from Machine Learning, Template Fitting, Hierarchical Bayesian Combination) and $z_{\mathrm{spec}}$. The top left plot compares the Machine Learning and Template Fitting $z_{\mathrm{phot}}$, the other three plots compare the $z_{\mathrm{phot}}$ predictions to $z_{\mathrm{spec}}$. Note that not all sources have a $z_{\mathrm{spec}}$, so there are many more points in the top left plot. The diagonal dashed line shows a one-to-one correspondence (if photo-$z$ predictions perfectly agreed with the spectroscopic redshifts).}
\label{fig:diagonal_plots_1}
\end{figure*}

\begin{figure*}
\includegraphics[scale=0.7]{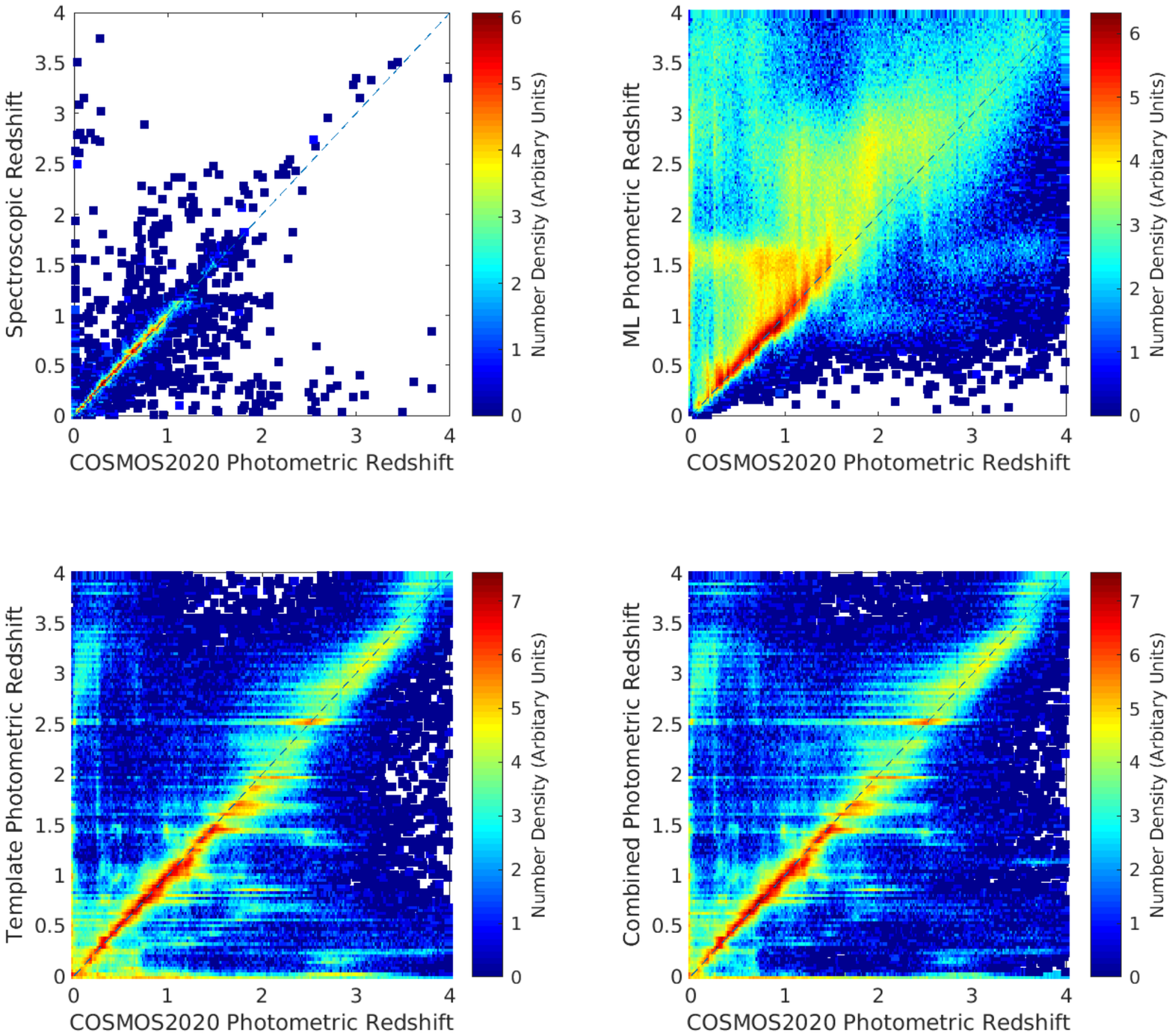}
\caption{Comparisons of the $z_{\mathrm{phot}}$ (from Machine Learning, Template Fitting, Hierarchical Bayesian Combination) and $z_{\mathrm{COSMOS2020}}$. The top left plot compares the spectroscopic redshifts and the COSMOS2020 redshifts, the other three plots compare the $z_{\mathrm{phot}}$ predictions to $z_{\mathrm{COSMOS2020}}$. Note that not all sources have a $z_{\mathrm{spec}}$, so there are many fewer points in the top left plot. The diagonal dashed line shows a one-to-one correspondence (if photo-$z$ predictions perfectly agreed with the spectroscopic redshifts)}
\label{fig:diagonal_plots_2}
\end{figure*}

Figure \ref{fig:diagonal_plots_1} shows the point estimates from the three photo-$z$ predictions (ML, template, Hierarchical Bayesian), and the spectroscopic redshifts. Figure \ref{fig:diagonal_plots_2} shows the point estimates from the three photo-$z$ predictions (ML, template, Hierarchical Bayesian), and the COSMOS2020 redshifts (and the spectroscopic redshifts to the COSMOS2020 redshifts). In Figure \ref{fig:diagonal_plots_1}, in the comparison of the ML and template fitting predictions, it can be seen the predictions agree for many sources (the data on the diagonal over $0<z<1$), but that there are many objects for which the predictions disagree (predominantly the fainter sources). For the photo-$z$ to spectroscopic redshift comparisons it can be seen that the photo-$z$ predictions are generally accurate for the spectroscopic sample for all three methods. The Hierarchical Bayesian predictions look qualitatively similar to the template fitting predictions, but with some outliers corrected (e.g. in Figure \ref{fig:diagonal_plots_1} the population at $z_{\mathrm{spec}}\sim0.5$ and $z_{\mathrm{phot}}\sim0.2$ for the template fitting is corrected for the Hierarchical Bayesian predictions). Similarly in Figure  \ref{fig:diagonal_plots_2} it can be seen that our photo-$z$ are in general agreement with the COSMOS2020 photo-$z$ out to $z\sim4$, although still with a moderate number of outliers (some of which may be due to inaccuracies in our redshifts, and others of which may be due to inaccuracies in COSMOS2020).

\begin{figure}
\includegraphics[scale=0.5]{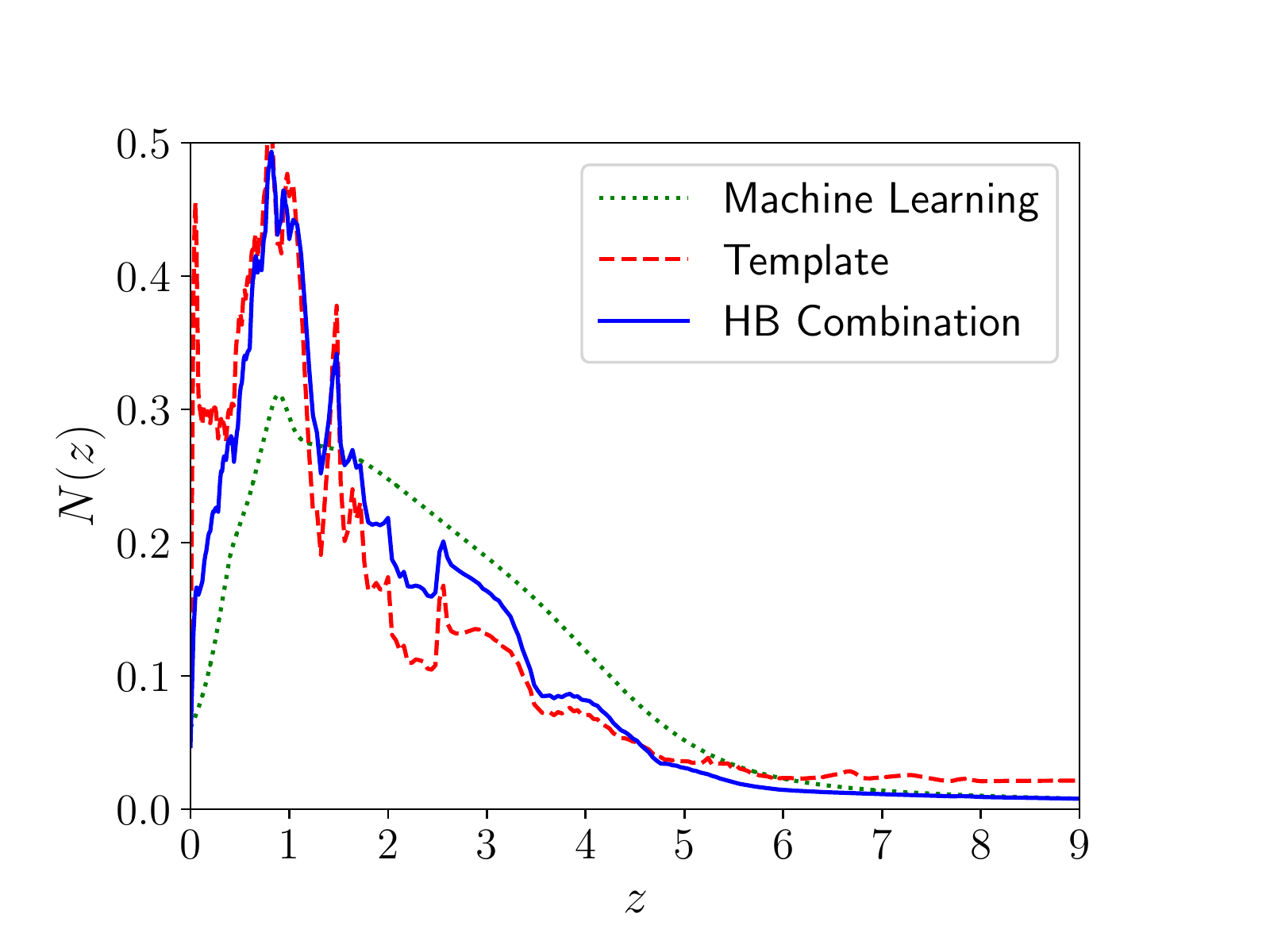}
\caption{The stacked pdfs from the ML, template, and Hierarchical Bayesian Combination photo-$z$ calculations.}
\label{fig:stacked_distribution}
\end{figure}

Figure \ref{fig:stacked_distribution} shows the normalised stacked pdf distributions, indicating the implied redshift distribution of our sample. We note that estimates of the population redshift distribution can also be derived with other Hierarchical Bayesian models \citet{Leistedt2016,Malz2020,Malz2021}. All three distributions are relatively similar and all peak at $z\sim1$, although there are some differences, in particular the sharp $z=0$ peak in the template-based distribution is not present in the HB Combination distribution, and the high redshift tails have different thicknesses.

\subsection{Metrics} \label{sec:metrics}

\begin{figure*}
\includegraphics[scale=0.8]{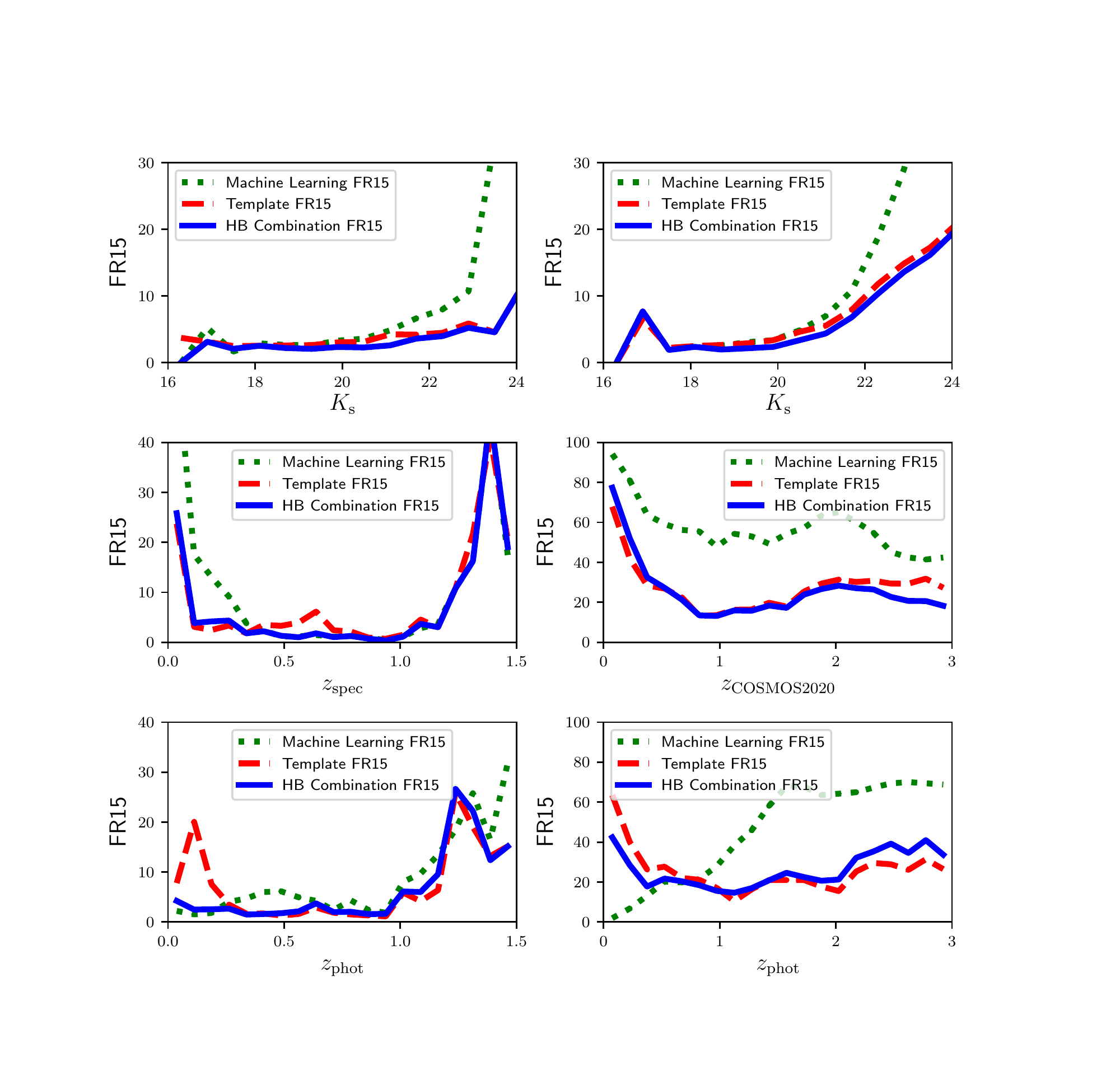}
\caption{The outlier fraction as a function of `true' redshift for our three photo-$z$ predictions. The first row shows as a function of $K_{\mathrm{s}}$-band magnitude, the second row shows as a function of spectroscopic or COSMOS2020 redshift, and the third row shows as a function of photometric redshift. In the left column the `true' redshifts are the spectroscopic redshift, in the right column the `true' redshifts are the COSMOS2020 redshifts (the columns are plotted with different redshift ranges, corresponding to the redshift ranges where there were sufficiently large numbers of galaxies). Note prediction performance is likely to be poorer for the sample as a whole as the spectroscopic sample isn't representative.}
\label{fig:fr15_plots}
\end{figure*}

Figure \ref{fig:fr15_plots} shows the outlier fraction FR15 (the fraction of sources with $|\Delta z|>0.15(1+z_{\mathrm{spec}})$) as compared to both spectroscopic redshift and COSMOS2020 redshift (for which $z_{\mathrm{COSMOS2020}}$ is used instead of $z_{\mathrm{spec}}$). The ML achieves a very low FR15 over $0.2<z<1.2$ when compared with the spectroscopic sample, but then degrades at higher and lower redshifts where there is less training data. The template FR15 is relatively flat up to $z\sim1.2$ and then also rises. The Hierarchical Bayesian photo-$z$ has similar but slightly better FR15 values to the template fitting predictions.

\begin{figure*}
\includegraphics[scale=0.8]{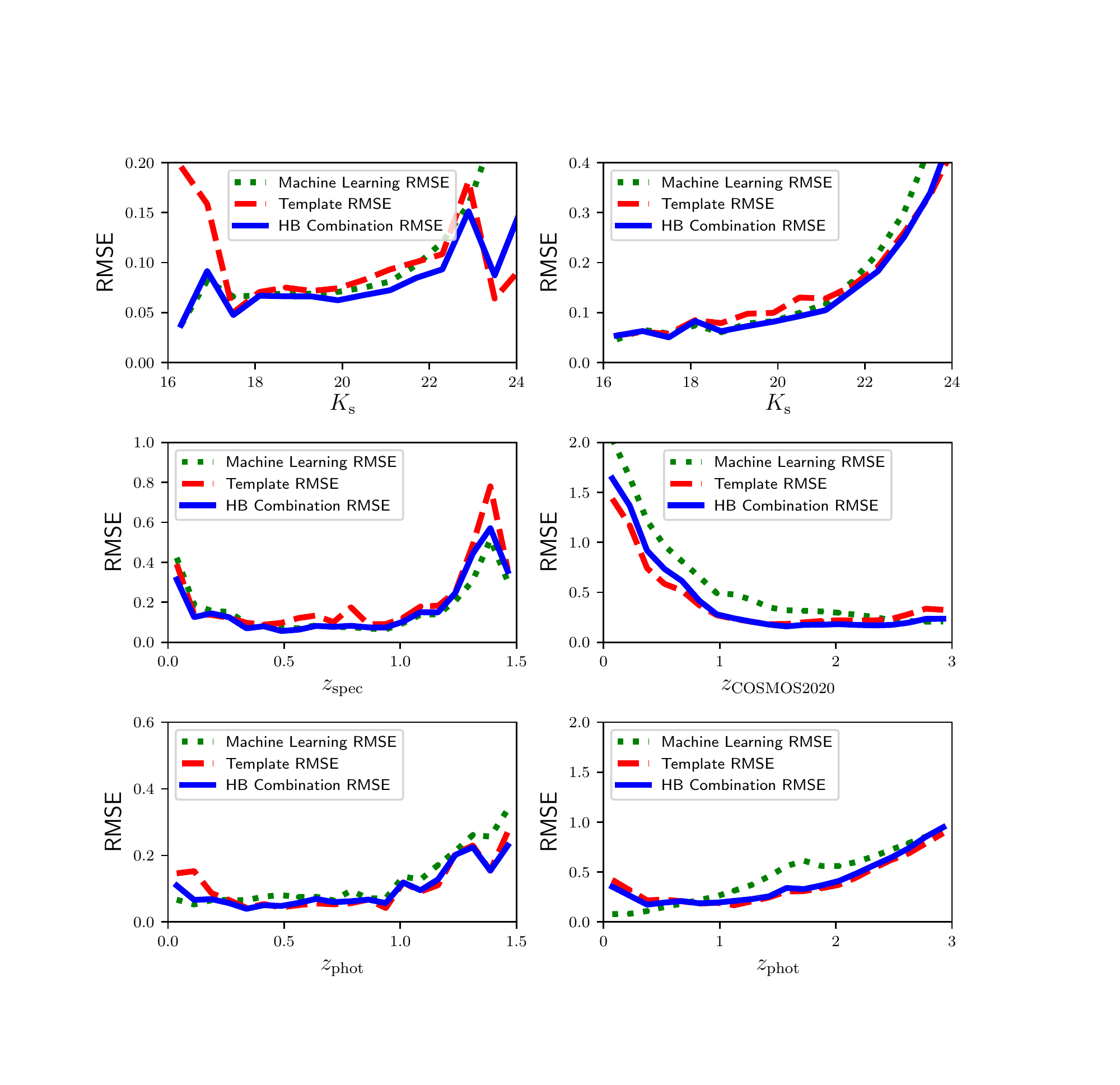}
\caption{The RMSE scatter as a function of `true' redshift for our three photo-$z$ predictions. The first row shows as a function of $K_{\mathrm{s}}$-band magnitude, the second row shows as a function of spectroscopic or COSMOS2020 redshift, and the third row shows as a function of photometric redshift. In the left column the `true' redshifts are the spectroscopic redshift, in the right column the `true' redshifts are the COSMOS2020 redshifts (the columns are plotted with different redshift ranges, corresponding to the redshift ranges where there were sufficiently large numbers of galaxies). Note prediction performance is likely to be poorer for the sample as a whole as the spectroscopic sample isn't representative.}
\label{fig:scatter_plots}
\end{figure*}

Figure \ref{fig:scatter_plots} shows the root mean square error (RMSE, $\sqrt{ \frac{1}{n}\Sigma_{i=1}^{n} (\frac{z_{\mathrm{spec}}-z_{\mathrm{phot}}}{1+z_{\mathrm{spec}}})^2 }$) scatter again as compared to both spectroscopic redshift and COSMOS2020 redshift (for which $z_{\mathrm{COSMOS2020}}$ is used instead of $z_{\mathrm{spec}}$). All methods have lowest RMSE over $0.2<z<1.2$ with the Hierarchical Bayesian model giving the lowest scatter for most redshifts.  Both the FR15 and RMSE are qualitatively similar to Figure 7 of \citet{Duncan2018b}.

\begin{figure*}
\includegraphics[scale=0.8]{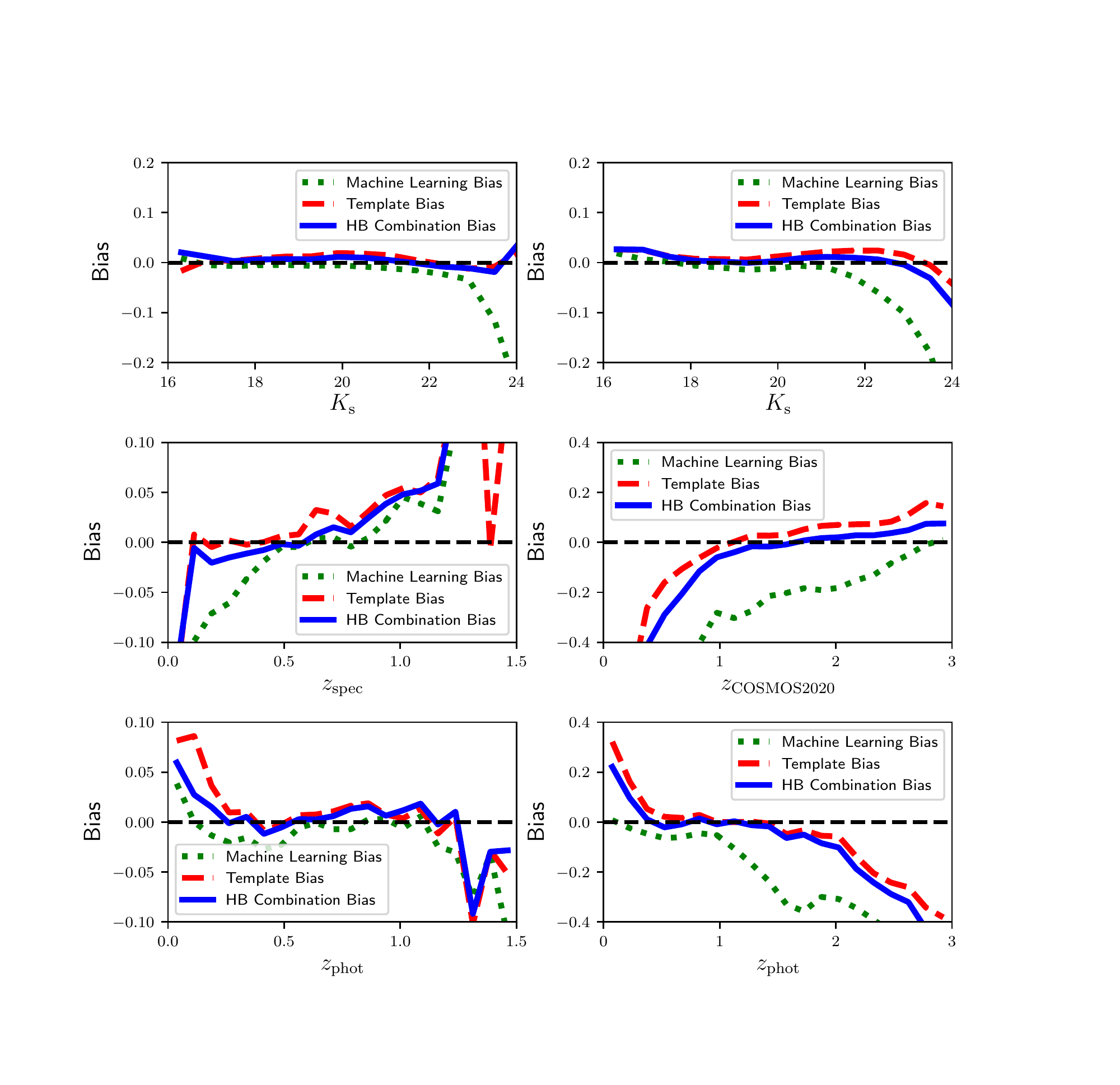}
\caption{The bias as a function of `true' redshift for our three photo-$z$ predictions. The first row shows as a function of $K_{\mathrm{s}}$-band magnitude, the second row shows as a function of spectroscopic or COSMOS2020 redshift, and the third row shows as a function of photometric redshift. In the left column the `true' redshifts are the spectroscopic redshift, in the right column the `true' redshifts are the COSMOS2020 redshifts (the columns are plotted with different redshift ranges, corresponding to the redshift ranges where there were sufficiently large numbers of galaxies). Note prediction performance is likely to be poorer for the sample as a whole as the spectroscopic sample isn't representative.}
\label{fig:bias_plots}
\end{figure*}

Figure \ref{fig:bias_plots} shows the Bias ($\frac{z_{\mathrm{spec}}-z_{\mathrm{phot}}}{1+z_{\mathrm{spec}}}$ when comparing to the spectroscopy, and $\frac{z_{\mathrm{COSMOS2020}}-z_{\mathrm{phot}}}{1+z_{\mathrm{COSMOS2020}}}$ when comparing to the COSMOS2020 redshifts). In most bins the Hierarchical combination has a value intermediate to the other two estimates.

In terms of $K_{\mathrm{s}}$-band magnitude dependence, it can be seen that for all three metrics, and all three estimates, performance is broadly better for brighter sources, with relatively flat quality out to $K_{\mathrm{s}}\approx 23$ for the ML predictions, and $K_{\mathrm{s}}\approx 24$ for the other two predictions - before rapid deterioration. The HB Combination is the highest performing of the three estimates for most magnitudes, both when compared to the spectroscopic sample, and to the COSMOS2020 sample. 

Figures \ref{fig:fr15_plots}, \ref{fig:scatter_plots} and \ref{fig:bias_plots} collectively show the key metrics (outlier fraction, root mean squared error, and bias respectively) for the performance of the three sets of photo-$z$ predictions, when compared to the spectroscopic and COSMOS2020 samples. Note that these data sets are not representative of the whole sample, so in general performance is likely to be poorer over the entire source population\footnote{Also the COSMOS2020 redshifts are not perfectly accurate themselves}. Furthermore the machine learning prediction is trained on the selfsame spectroscopic data, and thus should be expected to perform particularly well. In general the performance on the spectroscopic sample is very high on all three metrics. For the most part the Hierarchical Bayesian combination predictions outperforms the individual machine learning and template based results (although not quite for every single bin). Performance, in terms of consistency with COSMOS2020, for the entire sample was poorer than for the spectroscopic sample across the three metrics, although still high considering the depth of the sample. The Hierarchical Bayesian combination prediction was still broadly the best performing for most bins, but not nearly as consistently (although this is difficult to definitively draw conclusions about, as COSMOS2020 was also template-based, and thus might be expected to be methodologically correlated with our template fitting results). The performance of the three estimates are summarised in Table \ref{tab:summary}, where it can be seen that the Hierarchical Bayesian Combination performs best for all the comparisons with the spectroscopic data, and the template fitting best for two of the three metrics for the COSMOS2020 comparison, and Hierarchical Bayesian Combination best for the third (although only with marginal significance).

\begin{table*}
\scriptsize

\caption{Summary statistics of the three estimators (see also Figures \ref{fig:fr15_plots}, \ref{fig:scatter_plots} and \ref{fig:bias_plots}). The best result for each metric is in bold. Uncertainties are calculated with a bootstrapping method.}
\begin{tabular}{@{}cccc@{}}

\toprule
\textbf{Redshift Estimate}	& 	& \textbf{Metric}	&  \\
\midrule
& \textit{FR15}& \textit{RMSE}& \textit{Bias}\\
\midrule
\textit{Compared Against Spectroscopic Redshift}\\\midrule
Template Fitting  & {3.6$\pm$0.1} & {0.093$\pm$0.005}  & {0.0110$\pm$0.0006} \\
Machine Learning  & {4.8$\pm$0.2} & {0.089$\pm$0.002}  & {-0.0117$\pm$0.0007} \\
Hierarchical Combination  & {\textbf{2.8$\pm$0.1}} & {\textbf{0.077$\pm$0.003}}  & {\textbf{0.0044$\pm$0.0005}} \\
\midrule
\textit{Compared Against COSMOS2020 Redshift}\\\midrule
Template Fitting  & {24.7$\pm$0.05} & {\textbf{0.527$\pm$0.002}}  & {\textbf{-0.0773$\pm$0.0006}} \\
Machine Learning  & {56.2$\pm$0.07} & {0.788$\pm$0.001}  & {-0.388$\pm$0.001} \\
Hierarchical Combination  & {\textbf{24.6$\pm$0.05}} & {0.602$\pm$0.001}  & {-0.151$\pm$0.001} \\

\bottomrule
\end{tabular}\label{tab:summary}
\end{table*}

\begin{figure}
\includegraphics[scale=0.4]{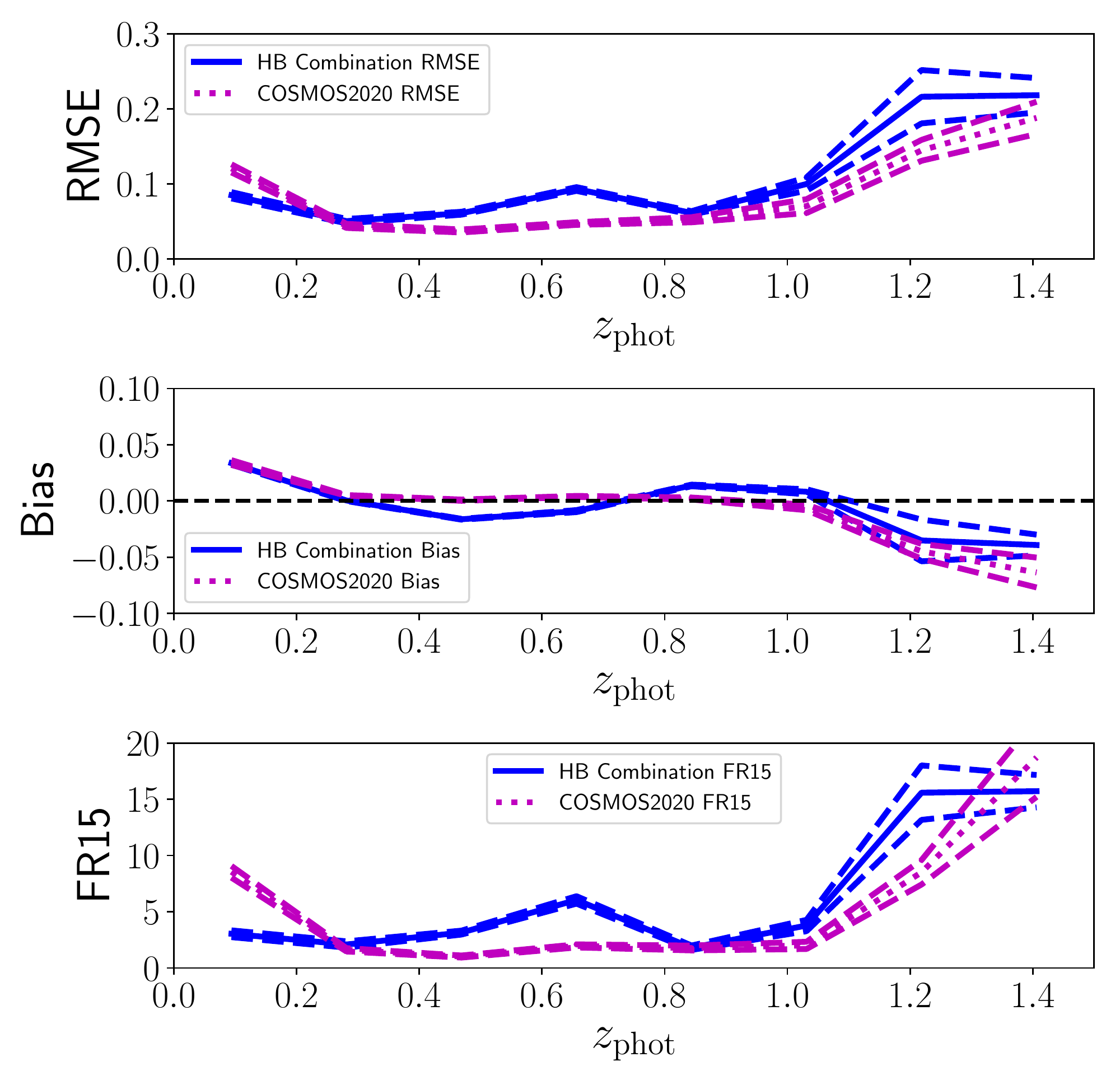}
\caption{A comparison of the performance of COSMOS2020 redshifts and our Hierarchical bayesian redshifts (when only the XMM-LSS spectroscopic redshifts are used in the training process), tested using the spectroscopic redshifts in the COSMOS field as the true redshifts. The top plot shows the RMSE scatter, the central plot shows the bias, and the bottom plot shows the FR15.  The dashed lines indicate the 1-$\sigma$ uncertainty on the measurements from a bootstrapping resampling analysis.}
\label{fig:COSMOS2015_comparison}
\end{figure}

If we wish to compare our final redshifts to the COSMOS2020 redshifts, we could make the comparison using the spectroscopic sample. However this would be slightly preferential to our redshifts as the spectroscopic redshifts in the COSMOS field were actually used in our training process. Thus we also calculate `XMM-LSS Trained' redshifts, where we only use the spectroscopic redshifts from the XMM-LSS field when training the ML model. We can then compare these results to the COSMOS2020 results\footnote{This of course slightly reduces the accuracy of our redshifts as they now have a smaller training set than our `main' redshift calculations.}. Figure \ref{fig:COSMOS2015_comparison} shows the RMSE, FR15 and bias for the COSMOS2020 redshifts and our `XMM-LSS Trained' Hierarchical Combination for the COSMOS spectroscopic sample. These results are summarised in Table \ref{tab:summary_xmm_only}. In particular note that a) the Template Fitting metrics are very similar to the corresponding values in Table \ref{tab:summary} (albeit not exactly the same because the test data in Table \ref{tab:summary_xmm_only} is just the COSMOS galaxies with spectroscopic redshifts, whereas Table \ref{tab:summary} is for spectroscopic redshifts from both COSMOS and XMM-LSS), and b) the `XMM-LSS Trained' Machine Learning metrics are poorer than our main Machine Learning estimate metrics, mainly because they have access to less training data.  It can be seen in Figure \ref{fig:COSMOS2015_comparison} that the two redshift predictions are relatively comparable - the COSMOS2020 redshift predictions have access to more photometric bands, but our redshift predictions have access to spectroscopic redshifts via the ML predictions. Our Hierarchical Bayesian Combinations are a little better at low-$z$ ($z<0.3$), and the COSMOS2020 predictions a little better at intermediate-$z$ ($0.4<z<0.8$). Table \ref{tab:summary_xmm_only} shows our calculations actually had very slightly better RMSE (due to higher performance at lower redshift where most of the galaxies are), and only slightly poorer bias and FR15.  Thus this work represents redshift estimations of comparable quality to COSMOS2020 (for bright and intermediate luminosity sources where evaluation is possible), now extended and homogeneous across both the COSMOS and XMM-LSS fields, calculated to fainter luminosities, and using fewer filters\footnote{Although as discussed, the quality for the parts of colour-magnitude space without spectroscopy is harder to validate.}. \textcolor{white}{XXX XXX XXX XXX XXX XXX XXX XXX XXX XXX XXX XXX XXX XXX XXX XXX XXX XXX XXX XXX XXX XXX XXX XXX XXX XXX XXX XXX XXX XXX XXX XXX XXX XXX XXX XXX XXX XXX XXX XXX XXX XXX XXX XXX XXX XXX XXX XXX XXX XXX XXX XXX XXX XXX XXX XXX XXX XXX XXX XXX XXX XXX XXX XXX XXX XXX XXX XXX XXX XX XXX XXX}

\begin{table*}
\scriptsize

\caption{Summary statistics of the COSMOS2020 estimates, the Template Fitting, the XMM-LSS-trained Machine Learning, and the XMM-LSS-trained Hierarchical Combination estimates (see also Figure \ref{fig:COSMOS2015_comparison}). The estimates are compared against the spectroscopic redshifts in the COSMOS field only. Uncertainties are calculated with a bootstrapping method. }
\begin{tabular}{@{}cccc@{}}
\toprule
\textbf{Redshift Estimate}	& 	& \textbf{Metric}	&  \\
\midrule
& \textit{FR15}& \textit{RMSE}& \textit{Bias}\\
\midrule
\textit{Compared Against Spectroscopic Redshift}\\\midrule
COSMOS2020  & {\textbf{2.9$\pm$0.1}} & {0.109$\pm$0.006}  & {\textbf{-0.0019$\pm$0.0007}} \\
Template Fitting  & {3.6$\pm$0.1} & {0.091$\pm$0.005}  & {0.0101$\pm$0.0007}  \\
XMM-LSS-trained Machine Learning  & {8.5$\pm$0.2} & {0.113$\pm$0.002}  & {-0.0295$\pm$0.0006} \\
XMM-LSS-trained Hierarchical Combination  & {3.9$\pm$0.1} & {\textbf{0.090$\pm$0.003}}  & {-0.0029$\pm$0.0006} \\

\bottomrule
\end{tabular}\label{tab:summary_xmm_only}
\end{table*}

\begin{figure}
\includegraphics[scale=0.5]{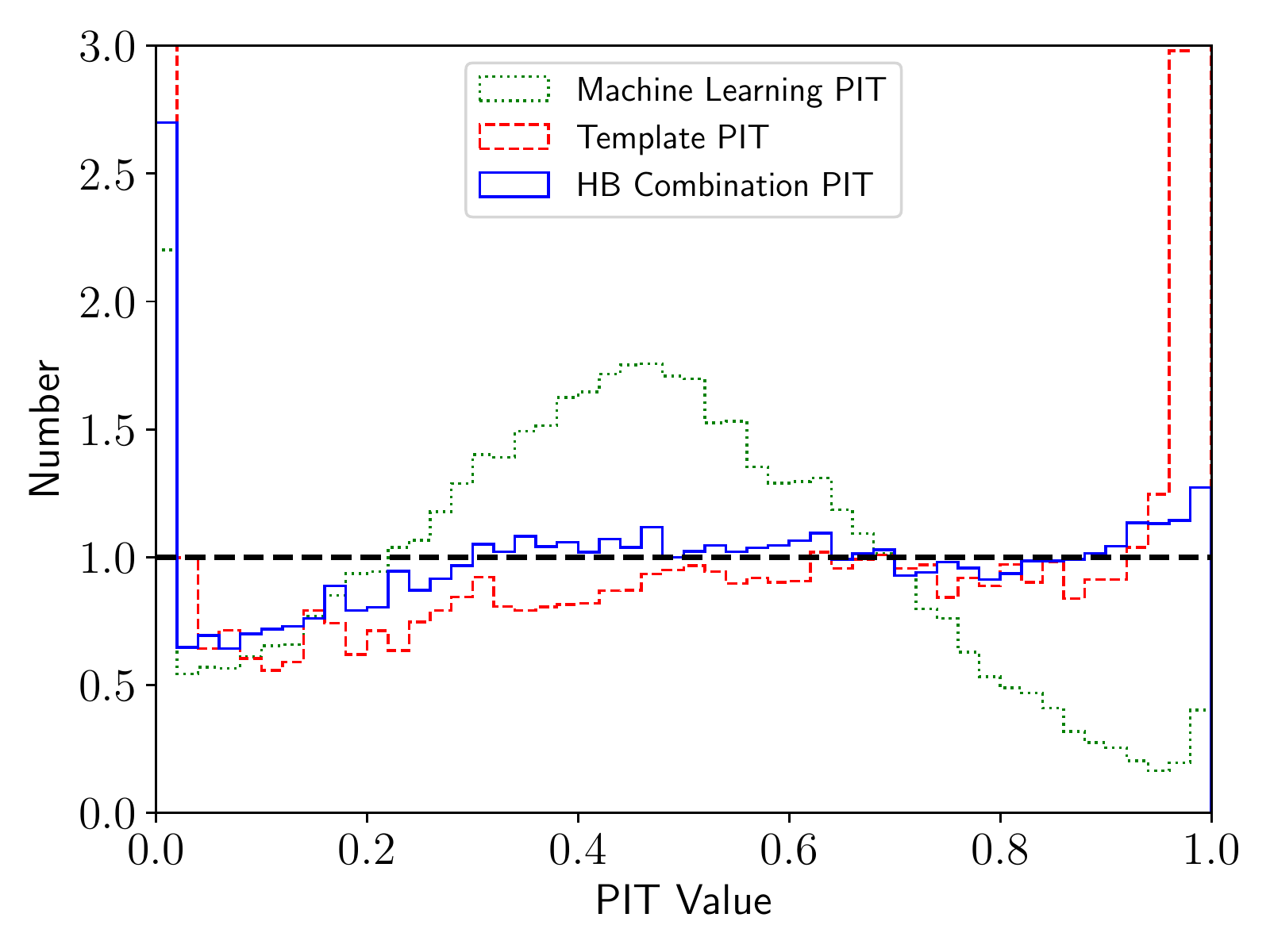}
\caption{A PIT plot for the GPz Machine Learning, Template Fitting and Hierarchical Bayesian Combination photo-$z$ pdfs.}
\label{fig:Q-Q}
\end{figure}

Figure \ref{fig:Q-Q} shows a Probability Integral Transform (PIT) plot for the pdfs of the spectroscopic sample (used extensively in the literature to assess photo-$z$ quality e.g. \citealp{Bordoloi2010}, see also Q-Q plots). PIT plots characterise the quality of the pdfs of the predictions, as opposed to the quality of the point estimates. To form a PIT  plot, first for each prediction the probability mass of the pdf less than the true value ($z_{\mathrm{spec}}$) is calculated. The PIT plot is a (often normalised) histogram of these values e.g. what is the distribution of probability mass in the pdfs less than the true value. A uniform distribution over (0,1) would correspond to a perfectly calibrated pdf e.g. 10\% of pdfs have 10\% of their probability mass less than the true value. In Figure \ref{fig:Q-Q} the Hierarchical Bayesian Model predictions are closest to the horizontal, indicating they are the most realistic pdfs. However deviations from the horizontal indicates that the pdfs are not calibrated quite perfectly\footnote{See for comparison the calibration process presented in \citet{Gomes2017}.}. Note that PIT plots only quantify the `realism' of pdfs, \textit{not} whether or not the predictions are useful or have any information content. This is most clearly illustrated by the trainZ estimator discussed in \citet{Schmidt2020}. This `algorithm' simply assigned every galaxy a pdf of the redshift distribution of the whole population. This assignment achieves a perfect PIT, but is terrible on almost all other metrics (e.g. point estimates like RMSE) as it contains no information content. As discussed earlier, our prior distribution used in the Hierarchical Bayesian combination mimics the population redshift distribution, so this contribution to the final pdfs will have high quality PIT. In particular for sources where both the template and the ML pdfs are deemed unreliable, the Hierarchical Bayesian combination will revert to this prior, and thus score highly on the PIT, but in general be a poor predictor. This is not necessarily either a positive or a negative property of the predictions, but it is important to emphasise here that PIT score alone is not an indication of quality of prediction. In any case the majority of sources have at least one reliable photo-$z$ estimate (see Figure \ref{fig:param_space}); only 19 sources had both $1/\beta_\mathrm{template}$ and $1/\beta_\mathrm{ML}$ less than 0.1. The prior still impacts the final pdf even if both estimates are reliable (e.g. if the estimates are `reliable', but with very large uncertainties), but the generally medium to high $1/\beta_i$ values, combined with the high performance on the point estimates, collectively implies that the pdfs are generally not being dominated by the prior.

\subsection{A possible $z\sim6.8$ galaxy?} \label{sec:high_z}

In this section we show how the combined Hierarchical Bayesian method could be used to check the validity of candidate high-$z$ galaxies, where although there are very few high-$z$ training spec-$z$, the higher accuracy of GPz at lower redshift may give more accurate weighting to low-$z$ solutions.

\begin{figure*}
\includegraphics[scale=0.9]{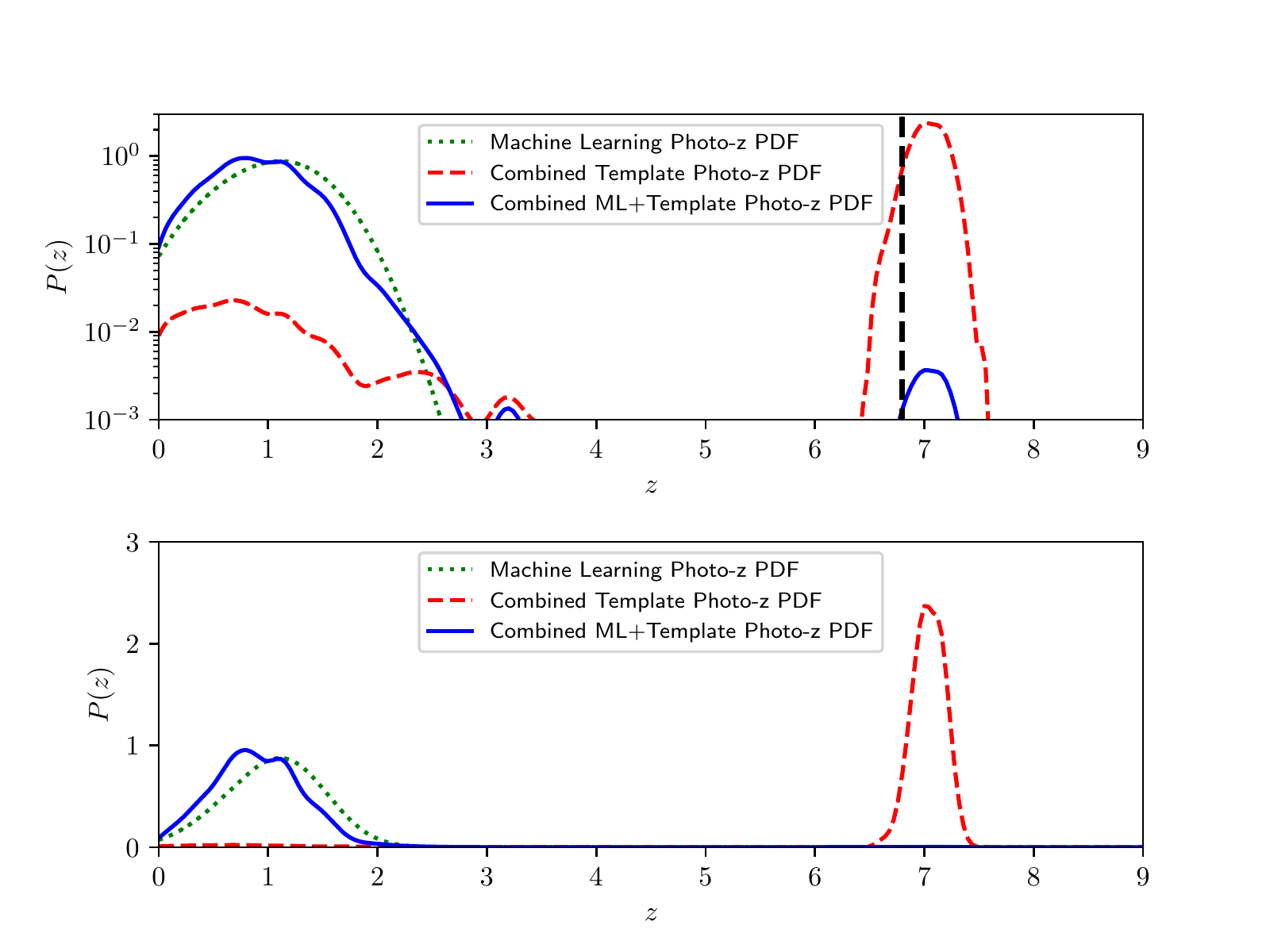}
\caption{The ML, template, and Hierarchical Bayesian pdfs for COS-87259 (top panel shows with a log-scale, the lower panel with a linear-scale). The vertical line shows the \citet{Endsley2021} redshift.}
\label{fig:high_z_pdf}
\end{figure*}

\citet{Endsley2021} report the detection of a possible $z\sim6.8$ massive star-forming galaxy. The source, COS-87259, was identified and assigned a redshift using the a Lyman-break narrow-band dropout technique - and is among the COSMOS sources we have analysed. The authors have many more bands (including narrow bands) than used in this work, but do note the possibility of a lower redshift interpretation. Our analysis assigned $z_{\mathrm{ML}}=1.04$, $z_{\mathrm{template}}=7.01$ and $z_{\mathrm{HB}}=0.78$ (pdfs shown in figure \ref{fig:high_z_pdf}). Our template fitting photo-$z$ thus was consistent with the dropout method, and the ML was consistent with a lower-$z$ solution. In agreement with \citet{Endsley2021} we found low $\chi^{2}$ values ($\chi^{2}_{\mathrm{galaxy}}=0.60$ and $\chi^{2}_{\mathrm{AGN}}=3.9$) indicating good fits, and that the template fitting solution should be reliable\footnote{Note however that we did not use any of the narrow bands used in \citet{Endsley2021}, so might expect to have a broader pdf than their results.}. However, for the ML $\nu/\sigma^2=0.013$, indicating that it wasn't excessively extrapolating, so that the ML prediction ought also be reliable. Furthermore, ML predictions do not rely on the set of templates used, which if incomplete may miss low-$z$ solutions. Both methods have similar $\beta_i$ weights for the combination of the pdfs. However the redshift prior distribution favours the low-$z$ solution to a large degree, meaning the $z_{\mathrm{HB}}$ is dominated by the $z_{\mathrm{ML}}$ value.

Which redshift value ought be believed? Firstly it should be noted that there are no training spectroscopic redshifts at $z\sim7$, so the ML will never predict a galaxy to be at $z\sim7$. The template fitting in this case gave a redshift estimation pdf with a $z\sim7$ peak, and a much smaller (by a factor of $\sim 100$ in probability) broad $z\sim1$ peak. Thus the question is whether the combination with the ML prediction and the prior via the HB model are correctly modifying the relative sizes of the peaks in the template fitting pdf. The final pdf is thus dependent on choice of weighting system (choice of $f_{\mathrm{bad}}$ etc.) and choice of prior. As \citet{Endsley2021} note, true confirmation of the redshift of the source will require spectroscopic follow-up. However the authors do present compelling evidence beyond simply the photometry (e.g. colocation on the sky with an over-density of other sources at $z\sim7$) that the source really is at high redshift. Assuming the source really is at high redshift, is our Hierarchical Bayesian model wrong to favour the lower redshift solution? We believe not necessarily, because COS-87259 was \textit{selected} based on the narrow-band photometry. It is perfectly consistent for the majority of sources with this broad-band colour-magnitude to be at lower-redshift (causing the ML prediction to take the lower-$z$ value), but a small fraction to be higher redshift. Which value to use depends on science goal and if false-positive or false-negative high-$z$ predictions are more costly. For context, taken at face value our $z_{\mathrm{template}}$ values would indicate 6 percent of our sources are at $z>6$, wheras our $z_{\mathrm{HB}}$ values would indicate closer to 1 percent. Finally, we would also note \citet{Endsley2021} identify radio continuum emission associated with COS-87259, which may further alter the appropriate redshift distribution prior.

\subsection{Choice of point estimate} \label{sec:mode_median}

\begin{figure}
\includegraphics[scale=0.4]{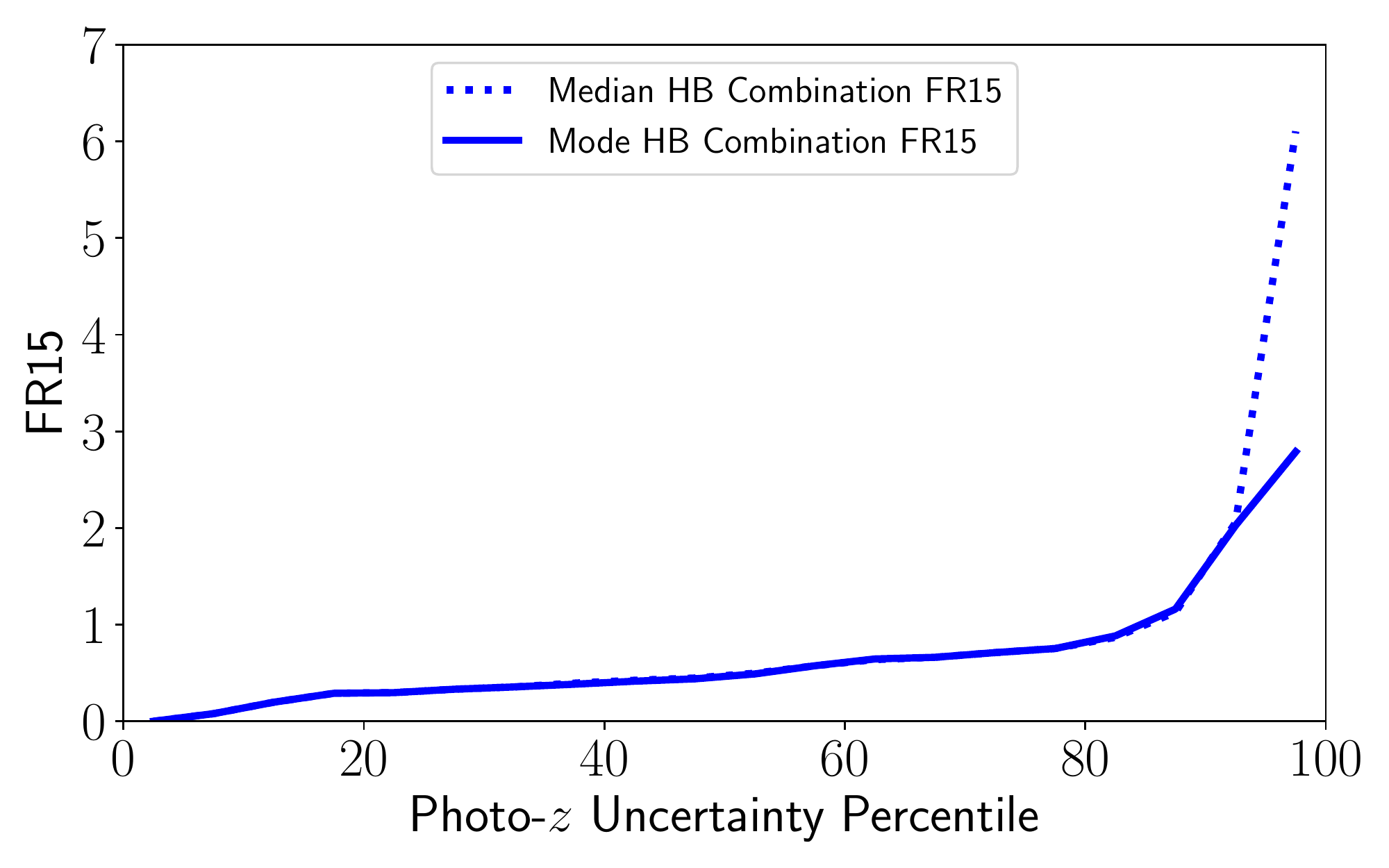}
\caption{The outlier fraction as a function of `true' redshift for the median and mode of our Hierarchical Bayesian photo-$z$ pdf predictions, compared to the spectroscopic redshifts. The data is plotted as a function of `Photo-$z$ Uncertainty Percentile' (as a function of percentile, ranked by uncertainty on estimate).}
\label{fig:median}
\end{figure}

As we mentioned in Section \ref{sec:best}, choice of point estimate depends on science goal. Figure \ref{fig:median} shows FR15 as a function of percentage of data for median and mode point estimates from the Hierarchical Bayesian estimates. It can be seen that for majority of sources the predictions are very similar, but that the mode slightly outperforms the median, mainly for the sources with the highest uncertainty. Everywhere else in this work the mode is used when a point estimate is used.

\subsection{Evaluation of Composite Template Fitting Estimates} \label{sec:agn_versus_galaxy}

\begin{figure}
\includegraphics[scale=0.4]{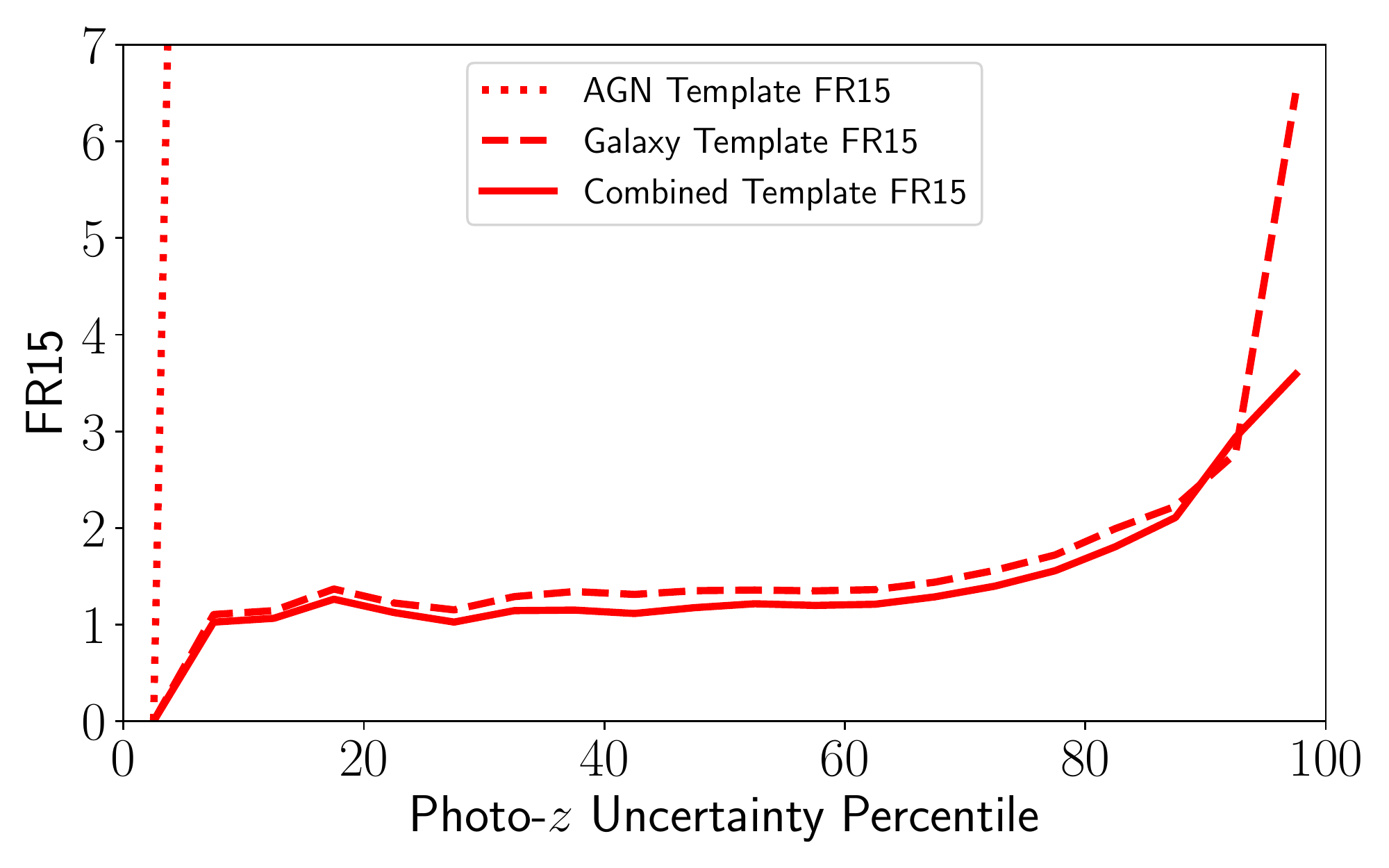}
\caption{The outlier fraction as a function of `true' redshift for our three template fitting based photo-$z$ predictions, compared to the spectroscopic redshifts. The data is plotted as a function of `Photo-$z$ Uncertainty Percentile' (as a function of percentile, ranked by uncertainty on estimate).}
\label{fig:agn_gal}
\end{figure}

We also test how great an improvement is achieved by including AGN template pdfs with the galaxy template pdfs, rather than just using galaxy templates (Section \ref{sec:gal_agn_combination_description}). Figure \ref{fig:agn_gal} shows FR15 as a function of percentage of data for the point estimates for the galaxy-only, AGN-only and Hierarchical Bayesian combination of the two (i.e. before the machine learning based predictions are added). It can be seen that the AGN-only prediction is much poorer than the other two estimators\footnote{Assuming one is interested in the quality of prediction across the whole sample. If ones science goal specifically concentrated on AGN then of course these predictions might be more helpful.}. The combined estimator is a little bit better than the galaxy-only prediction for most of the data, and much better for the final $\sim10$ percent of the data where the uncertainties are greatest (likely the AGN in the sample).

\section{Conclusions} \label{sec:conclusions}

We have presented the calculation of photometric redshift estimates for sources with deep optical and near infrared data over the COSMOS and XMM-LSS fields. We calculate template fitting based photometric redshifts for both galaxy and AGN templates using {\sc LePhare} (building on \citealp{Adams2020}), as well as machine learning based photometric redshifts using GPz (building on \citealp{Hatfield2020a}), and then used a Hierarchical Bayesian model to combine them (using the method of \citealp{Duncan2018b}). By combining template fitting and machine learning we achieve predictions that take the best aspects of both approaches to photo-$z$ calculation. These redshifts were then tested by comparison to the \citet{Weaver2021} COSMOS2020 photometric redshifts, which are still photometric redshifts, but had access to a higher number of bands. Our redshifts are of comparable quality to the COSMOS2020 redshifts - the information from the spectroscopic training set can make up some of the loss of information from having fewer bands.

The redshifts calculated in this work thus represent the most accurate set of redshifts for a catalogue this large of deep multi-wavelength photometry over multi-square degree surveys. Calculating photometric redshifts is a key challenge in extragalactic astronomy - this work and the resulting dataset represents an important large set of reliable high-quality photo-$z$ for future science use over these key extragalactic fields.

\section*{Acknowledgements}

PH acknowledges generous support from the Hintze Family Charitable Foundation through the Oxford Hintze Centre for Astrophysical Surveys, and acknowledges travel support provided by STFC for UK participation in Rubin through grant ST/N002512/1. RB acknowledges support from an STFC Ernest Rutherford Fellowship [grant number ST/T003596/1]. This publication arises from research funded by the John Fell Oxford University Press Research Fund.

Based on data products from observations made with ESO Telescopes at the La Silla or Paranal Observatories under ESO programme ID 179.A- 2006. Based on observations obtained with MegaPrime/MegaCam, a joint project of CFHT and CEA/IRFU, at the Canada-France-Hawaii Telescope (CFHT) which is operated by the National Research Council (NRC) of Canada, the Institut National des Science de l'Univers of the Centre National de la Recherche Scientifique (CNRS) of France, and the University of Hawaii. This work is based in part on data products produced at Terapix available at the Canadian Astronomy Data Centre as part of the Canada-France-Hawaii Telescope Legacy Survey, a collaborative project of NRC and CNRS.

The Hyper Suprime-Cam (HSC) collaboration includes the astronomical communities of Japan and Taiwan, and Princeton University. The HSC instrumentation and software were developed by the National Astronomical Observatory of Japan (NAOJ), the Kavli Institute for the Physics and Mathematics of the Universe (Kavli IPMU), the University of Tokyo, the High Energy Accelerator Research Organization (KEK), the Academia Sinica Institute for Astronomy and Astrophysics in Taiwan (ASIAA), and Princeton University. Funding was contributed by the FIRST program from Japanese Cabinet Office, the Ministry of Education, Culture, Sports, Science and Technology (MEXT), the Japan Society for the Promotion of Science (JSPS), Japan Science and Technology Agency (JST), the Toray Science Foundation, NAOJ, Kavli IPMU, KEK, ASIAA, and Princeton University.

This paper makes use of software developed for the Large Synoptic Survey Telescope. We thank the LSST Project for making their code available as free software at \url{http://dm.lsst.org}.

This paper is based, in part, on data collected at the Subaru Telescope and retrieved from the HSC data archive system, which is operated by Subaru Telescope and Astronomy Data Center at National Astronomical Observatory of Japan. Data analysis was in part carried out with the cooperation of Center for Computational Astrophysics, National Astronomical Observatory of Japan.

\section*{Data Availability}

The photo-$z$ in this study will be made available online at time of publication. Other derived data generated in this research will be shared on reasonable request to the corresponding author.

\bibliographystyle{mn2e_mod}
\bibliography{VIDEO_photoz}

\bsp

\label{lastpage}

\end{document}